\documentclass[  superscriptaddress, amsmath,amssymb, aps,twocolumn,longbibliography]{revtex4-1} 
\usepackage{graphicx}
\usepackage{dcolumn}
\usepackage{amsthm}
\usepackage{bm}

\pdfoutput=1
\usepackage{graphicx,graphics,epsfig,subfigure,times,bm,bbm,amssymb,amsmath,amsthm,mathrsfs,MnSymbol}
\usepackage{gensymb}
\usepackage{amsfonts}
\usepackage[matrix,frame,arrow]{xypic}
\usepackage[pdfstartview=FitH]{hyperref}
\hypersetup{
    colorlinks=true,       
    linkcolor=red,          
    citecolor=magenta,        
    filecolor=magenta,      
    urlcolor=cyan,           
    runcolor=cyan}
\usepackage{epstopdf}
\usepackage[pdftex]{color}
\usepackage{braket}
\usepackage{enumerate}
\usepackage[normalem]{ulem}
\usepackage[usenames,dvipsnames]{xcolor}
\usepackage{multirow}
\usepackage{mathtools}
\usepackage{makecell}

\definecolor{orange}{rgb}{1,0.5,0}

\begin{document}

\title{Variational generation of spin squeezing on one-dimensional quantum devices 
\\ with nearest-neighbor interactions}

\author{Zheng-Hang Sun}
\affiliation{Institute of Physics, Chinese Academy of Sciences, Beijing 100190, China}
\affiliation{School of Physical Sciences and CAS Center of Excellence for Topological Quantum Computation, University of Chinese Academy of Sciences, Beijing 100190, China}

\author{Yong-Yi Wang}
\affiliation{Institute of Physics, Chinese Academy of Sciences, Beijing 100190, China}
\affiliation{School of Physical Sciences and CAS Center of Excellence for Topological Quantum Computation, University of Chinese Academy of Sciences, Beijing 100190, China}

\author{Yu-Ran Zhang}
\email{yuranzhang@scut.edu.cn}
\affiliation{School of Physics and Optoelectronics, South China University of Technology, Guangzhou 510640, China}
\affiliation{Theoretical Quantum Physics Laboratory, Cluster for Pioneering Research, RIKEN, Wakoshi, Saitama 351-0198, Japan}
\affiliation{Center for Quantum Computing, RIKEN, Wakoshi, Saitama 351-0198, Japan}

\author{Franco Nori}
\email{fnori@riken.jp}
\affiliation{Theoretical Quantum Physics Laboratory, Cluster for Pioneering Research, RIKEN, Wakoshi, Saitama 351-0198, Japan}
\affiliation{Center for Quantum Computing, RIKEN, Wakoshi, Saitama 351-0198, Japan}
\affiliation{Physics Department, University of Michigan, Ann Arbor, MI 48109-1040, USA}

\author{Heng Fan}
\email{hfan@iphy.ac.cn}
\affiliation{Institute of Physics, Chinese Academy of Sciences, Beijing 100190, China}
\affiliation{School of Physical Sciences and CAS Center of Excellence for Topological Quantum Computation, University of Chinese Academy of Sciences, Beijing 100190, China}
\affiliation{Songshan Lake  Materials Laboratory, Dongguan 523808, Guangdong, China}
\affiliation{Beijing Academy of Quantum Information Sciences, Beijing 100193, China}
\affiliation{Hefei National Laboratory, Hefei 230088, China}

\begin{abstract}
\noindent Efficient preparation of spin-squeezed states is important for quantum-enhanced metrology. Current protocols for generating strong spin squeezing rely on either high dimensionality or long-range interactions. A key challenge is how to generate considerable spin squeezing in one-dimensional systems with only nearest-neighbor interactions. 
Here, we develop variational spin-squeezing algorithms to solve this problem. We consider both digital and analog quantum circuits for these variational algorithms. After the closed optimization loop of the variational spin-squeezing algorithms, the generated squeezing can be comparable to the strongest squeezing created from two-axis twisting. By analyzing the experimental imperfections, the variational spin-squeezing algorithms proposed in this work are feasible in recent developed noisy intermediate-scale quantum computers.  
\end{abstract}
\pacs{Valid PACS appear here}
\maketitle

\section{INTRODUCTION}

Spin squeezing plays a central role in the field of quantum metrology and the characterization of quantum entanglement~\cite{RevModPhys.90.035005,MA201189}. Generating highly spin-squeezed states is of fundamental interest because of its promising applications in high-precision measurements~\cite{qm1,qm2,qm3,PhysRevLett.105.053601}, such as the search for dark matter axions~\cite{Backes:2021wd}. Different from the Greenberger-Horne-Zeilinger state, which is very fragile to noise~\cite{PhysRevLett.79.3865}, spin-squeezed states are more robust against amplitude-damping noise~\cite{MA201189,PhysRevA.81.022106}. A paradigmatic protocol for generating  spin-squeezed states is one-axis twisting (OAT)~\cite{PhysRevA.47.5138}, which can be straightforwardly realized in several experimental platforms with long-range interactions, including atomic Bose-Einstein condensates~\cite{qm2,qm3,PhysRevLett.98.200405,PhysRevA.92.023603,qm4,qm5}, trapped ions~\cite{ti_oat}, Rydberg-dressed atoms~\cite{2023arXiv230308078E}, and superconducting circuits with all-to-all connectivity~\cite{qc_all_1,qc_all_2,PhysRevLett.128.150501}. 

Despite the enormous progress in generating squeezing via OAT, there are two key challenges in this field. The first one is that OAT relies on the infinite-range interactions, while only a few experimental platforms can achieve  sufficiently long-range or infinite-range interactions.  Consequently, exploring the possibility for generating highly spin-squeezed states in systems with short-range interactions is highly desirable. The second challenge is the realization of two-axis twisting (TAT), generating stronger squeezing than OAT~\cite{PhysRevA.47.5138}. Even though several schemes for achieving TAT have been proposed~\cite{tat_1,tat_2,tat_3,tat_4,tat_5}, its experimental realization remains absent due to the limited controllability and flexibility of artificial quantum systems.  

Recent studies have shown that spin squeezing comparable to the best one generated from OAT can be achieved in two- and three-dimensional short-range systems~\cite{PhysRevLett.125.223401,2023arXiv230308053B,2023arXiv230109636B}. It indicates that higher-dimensional systems have the potential to yield stronger squeezing~\cite{PhysRevLett.112.103601,vqe_sss1}. Nevertheless, there remains a more challenging problem, i.e., whether \emph{one-dimensional} (1D) quantum devices with only \emph{nearest-neighbor} interactions, as one of the most experimentally achievable systems in quantum simulation, can generate highly spin-squeezed states that are comparable to those from OAT, or even from TAT.

In the past few years, tremendous developments on variational quantum algorithms~\cite{vq_1,vq_2,vq_3,vq_4,vq_5} opened new avenues for improving the performance of quantum sensors~\cite{vqe_sss1,vqe_sss2,vqe_sss3,vqe_sss4,vqe_sss5,vqe_sss6}, providing a method for overcoming the aforementioned challenges. Here, by adopting the squeezing parameter~\cite{PhysRevLett.122.090503} (lower value of the parameter corresponding to a better spin-squeezed state) as the objective function, we propose variational spin-squeezing algorithms on 1D quantum devices with nearest-neighbor interactions to achieve not only OAT but also TAT.

\begin{figure}[h!]
	\centering
	\includegraphics[width=1\linewidth]{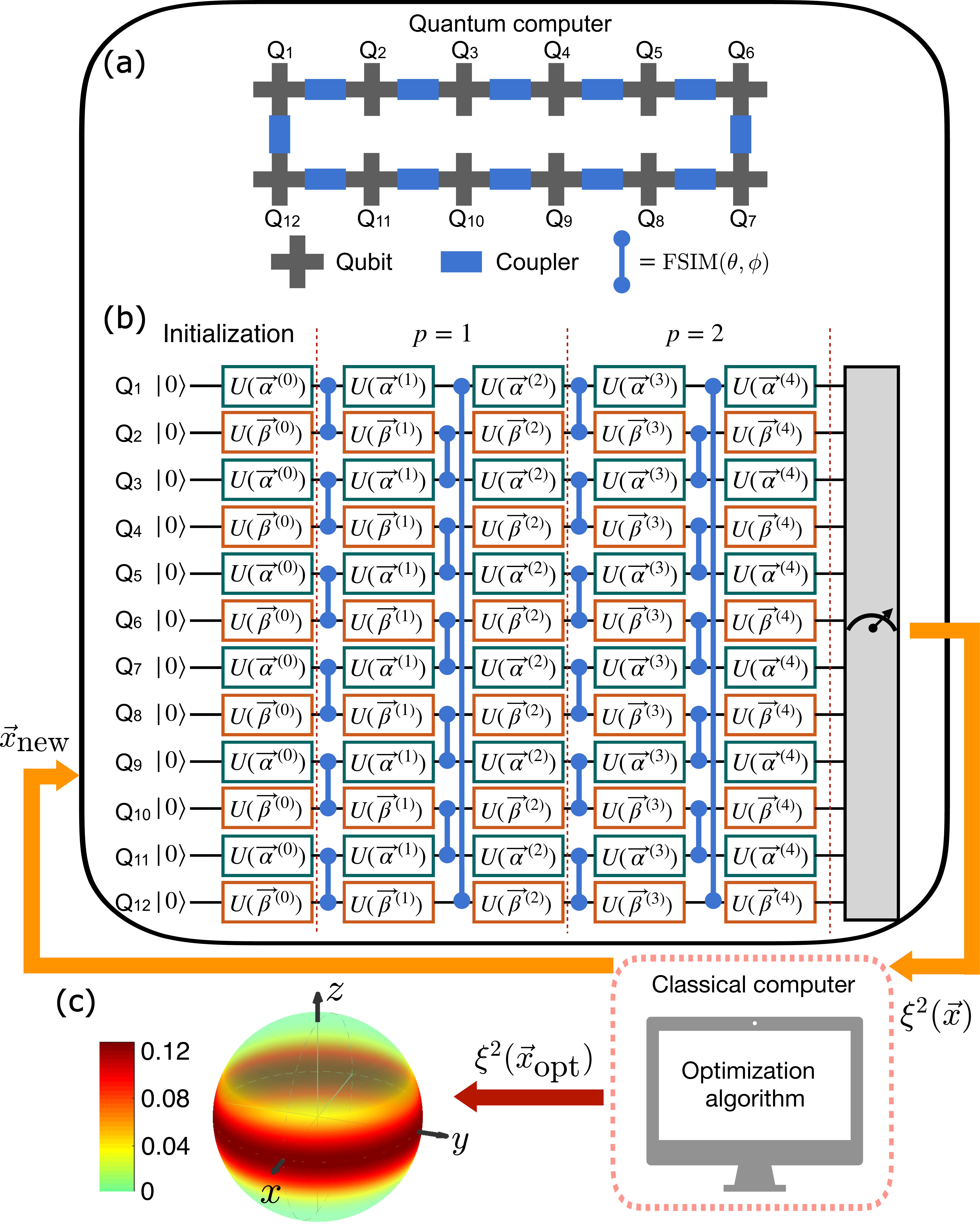}\\
	\caption{ (a) A digital quantum computer consisting of a loop of superconducting qubits connected with couplers. (b) Schematic of the parameterized quantum circuit (PQC) based on the alternating layered ansatz with $N=12$ qubits and the number of layers $p=2$. Here, the two-qubit gate notation refers to the Fermionic Simulation (FSIM) gate defined by Eq.~(\ref{fsim}). The initial state is chosen as $|\psi_{0}\rangle = |00...0\rangle$, where $|0\rangle$ is the  eigenstate of $\hat{\sigma}^{z}$ with the eigenvalue $-1$. The variational parameters are denoted as $\vec{x}$. One layer of the PQC contains two layers of entangler $\hat{U}_{E,1}$ and $\hat{U}_{E,2}$, each of which is followed by a layer of single-qubit rotations. After measuring the final state of the PQC, the value of objective function $\xi^{2}(\vec{x})$ can be obtained and then is fed into the classical computer, where the optimization algorithm is running, finding the next set of variational parameters $\vec{x}_{\text{new}}$. Then, $\vec{x}_{\text{new}}$ is fed into the quantum computer, completing one loop of the algorithm. Finally, by iteratively running several loops, one can obtain the optimized objective function $\xi^{2}_{\text{opt}} = \xi^{2}(\vec{x}_{\text{opt}})$. (c) The Husimi $Q$ function for the optimized spin-squeezed states generated by the 12-qubit PQC with $p=3$.}\label{fig1}
\end{figure}

The performance of variational quantum algorithms greatly depends on the choice of ansatz for designing the parameterized quantum circuit (PQC)~\cite{vq_5}.
Here, we first focus on a PQC based on the alternating layered ansatz (ALA), which can be realized in several platforms of digital quantum computers, ranging from near-term superconducting circuits~\cite{sc1,sc2,sc3,sc4,sc7,sc8} [also see Fig.~\ref{fig1}(a) and (b)] to neutral Rydberg atoms~\cite{rydberg_gate1,rydberg_gate2}. Flexible digital quantum computers allow us to implement gate-model-based circuits using different ansatzes. We choose the ALA because of its deterministic expressibility~\cite{Nakaji2021expressibilityof} and ameliorated barren-plateau phenomena~\cite{tat_expr1}. Then, we consider a hardware-efficient PQC that is suitable for analog quantum computers, where all qubits in the circuit can be entangled via the global unitary evolution of a tailored Hamiltonian. We numerically study the performance of the aforementioned two variational spin-squeezing algorithms. Remarkably, we find that a spin-squeezed state with a squeezing parameter comparable to the lowest one obtained from TAT can be generated from both the shallow-depth optimized digital and analog PQCs. We also analyze the influence of experimental imperfections, and explore the methods to reduce errors in the quantum circuit of the variational spin-squeezing algorithms.




\begin{figure*}
	\centering
	\includegraphics[width=1\linewidth]{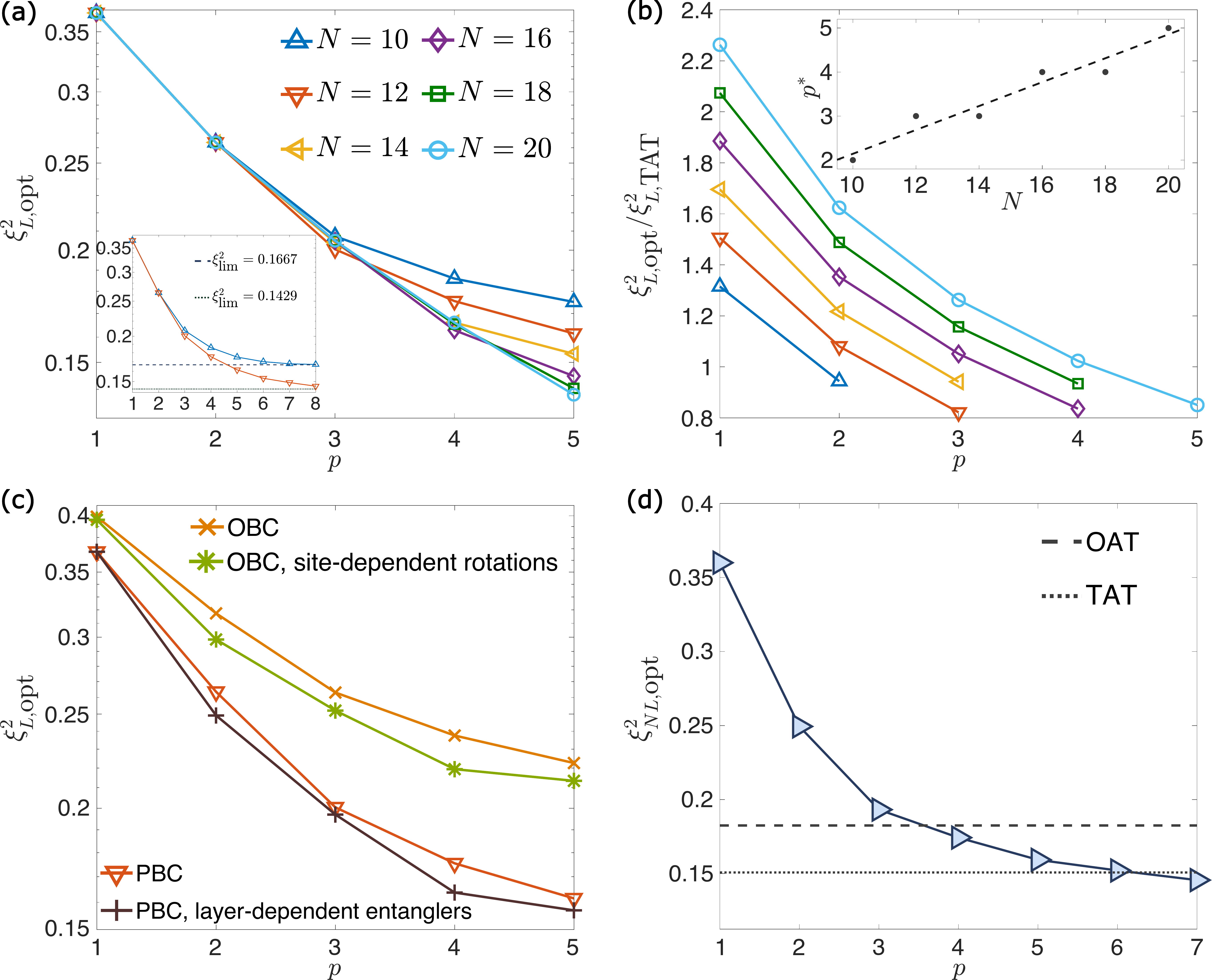}\\
	\caption{ (a) The optimized linear Ramsey squeezing parameter $\xi^{2}_{L,\text{opt}}$ as a function of the depth $p$ of the PQC for different numbers of qubits $N=10$, $12$, $14$, $16$, $18$, and $20$. The inset of (a) shows the $\xi^{2}_{L,\textrm{opt}}$ for the deeper depth of the PQC $p$ with $N=10$ and $N=12$. The dashed and dotted horizontal lines represent the value of the fundamental limit of the linear Ramsey squeezing parameter with $N=10$ and $N=12$, respectively.  (b) The ratio of $\xi^{2}_{L,\text{opt}}$ to the lowest squeezing parameter $\xi^{2}_{L,\text{TAT}}$ obtained from TAT, i.e., $\xi^{2}_{L,\text{opt}}/\xi^{2}_{L,\text{TAT}}$, as a function of the depth $p$. For each system size $N$, only the data with $p\leq p^{*}$ are plotted, where $p^{*}$ denotes the smallest depth required for achieving $\xi^{2}_{L,\text{opt}}/\xi^{2}_{L,\text{TAT}}<1$. The inset of (b) shows the $p^{*}$ as a function of $N$, where the dashed line is the linear fitting. (c) The optimized squeezing parameter $\xi^{2}_{L,\text{opt}}$ as a function of the depth $p$ for different designs of the PQC with $N=12$. (d) The optimized nonlinear Ramsey squeezing parameter $\xi^{2}_{NL,\text{opt}}$ as a function of the depth $p$ for $N=10$ qubits. The dashed and dotted lines in (d) represent the lowest nonlinear Ramsey squeezing parameter obtained from OAT and TAT, respectively. Note that (a) and (c) have a vertical logarithmic axis. }\label{fig2}
\end{figure*}


\section{Variational spin-squeezing algorithm with digital quantum circuits}

The workflow of variational spin-squeezing algorithms consists of four parts (see Fig.~\ref{fig1}): (i) the objective function; (ii) the PQC; (iii) measurement of the squeezing parameter; and (iv) the classical optimizer. Here, for an $N$-qubit system, the spin squeezing parameter, chosen as the objective function, is defined as~\cite{PhysRevLett.122.090503}
\begin{eqnarray}
\xi^{2}[\hat{\rho},\hat{\mathbf{S}}] = \frac{N}{\lambda_{\text{max}}(\mathbb{M}[\hat{\rho},\hat{\mathbf{S}}])},
\label{ssp}
\end{eqnarray}
where $\hat{\mathbf{S}}$ is a family of accessible operators, and $\lambda_{\text{max}}$ is the maximum eigenvalue of the matrix $\mathbb{M}[\rho,\hat{\mathbf{S}}]$. For the linear Ramsey squeezing parameter $\xi^{2}_{L}$, $\hat{\mathbf{S}} = \hat{\mathbf{S}}_{L} = (\hat{J}_{x},\hat{J}_{y},\hat{J}_{z})$ ($\hat{J}_{\alpha} = \frac{1}{2}\sum_{i=1}^{N}\hat{\sigma}_{\alpha}^{i}$, $\alpha\in\{x,y,z\}$), and 
\begin{eqnarray}
\mathbb{M}[\hat{\rho},\hat{\mathbf{S}}] = \mathbb{C}^{T}[\hat{\rho},\hat{\mathbf{S}}] \ \mathbb{V}^{-1}[\hat{\rho},\hat{\mathbf{S}}] \ \mathbb{C}[\hat{\rho},\hat{\mathbf{S}}],
\label{m}
\end{eqnarray}
with
\begin{eqnarray}
\mathbb{V}_{ij}[\hat{\rho},\hat{\mathbf{S}}] = \langle\{\hat{\mathbf{S}}_{i},\hat{\mathbf{S}}_{j}\}\rangle/2 - \langle\hat{\mathbf{S}}_{i}\rangle\langle\hat{\mathbf{S}}_{j}\rangle  
\label{v}
\end{eqnarray}
and 
\begin{eqnarray}
\mathbb{C}[\hat{\rho},\hat{\mathbf{S}}]=-i\langle[\hat{\mathbf{S}}_{i},\hat{\mathbf{S}}_{j}]\rangle. 
\label{c}
\end{eqnarray}
The parameter $\xi^{2}_{L}$ can be easily measured in near-term quantum computers~\cite{PhysRevLett.128.150501}, as discussed below in Appendix A. 
The definition of the squeezing parameter (\ref{ssp}) can be generalized to higher-order nonlinear squeezing parameter, which characterizes the entanglement of non-Gaussian states. 

For digital quantum computers, we design  the PQC enlightened by the ALA. As shown in Fig.~\ref{fig1}(b), the circuit is comprised of arbitrary single-qubit rotations with the angles $\vec{\omega}^{(k)}  \equiv (\omega_{1}^{(k)}, \omega_{2}^{(k)}, \omega_{3}^{(k)})$, i.e., 
\begin{equation}
\hat{U}_{j}(\vec{\omega}^{(k)}) = e^{-i\hat{\sigma}_{j}^{z}\omega_{1}^{(k)}}e^{-i\hat{\sigma}_{j}^{x}\omega_{2}^{(k)}}e^{-i\hat{\sigma}_{j}^{z}\omega_{3}^{(k)}},\label{single_1}
\end{equation}
and two kinds of layers of Fermionic Simulation (FSIM) gates 
\begin{equation}
\hat{U}_{E,1} = \bigotimes_{i=1}^{N/2}\text{FSIM}(\theta,\phi)_{2i-1,2i}
\label{ue1}
\end{equation}
and 
\begin{equation}
\hat{U}_{E,2} = \bigotimes_{i=1}^{N/2}\text{FSIM}(\theta,\phi)_{2i,2i+1}
\label{ue2}
\end{equation}
 being the entanglers, where $\text{FSIM}(\theta,\phi)_{j,j+1}$ refers to the FSIM gate applied on the qubit pair consisting of the $j$-th and $(j+1)$-th qubit with
\begin{equation}
\text{FSIM}(\theta,\phi) =
{\left( \begin{array}{cccc}
1 & 0 & 0 & 0 \\
0 & \cos\theta & -i\sin\theta & 0 \\
0 & -i\sin\theta & \cos\theta & 0 \\
0 & 0 & 0 & e^{-i\phi}
\end{array}
\right ) }.\label{fsim}
\end{equation}
The FSIM gate in (\ref{fsim}) plays a key role in the simulation of electronic structures~\cite{PhysRevLett.120.110501,sc3}, and can provide sufficient entanglement for demonstrating quantum advantage~\cite{sc1,sc2}. More importantly, a continuous set of high-fidelity FSIM gates with different control angles $\theta$ and $\phi$ can be experimentally realized~\cite{PhysRevLett.125.120504}.

The adjustable couplers in Fig.~\ref{fig1}(a) enable us to individually implement the layer of entangler $\hat{U}_{E,1}$ and $\hat{U}_{E,2}$. Here, for simplicity, we choose the same value of the angles $(\theta, \phi)$ in FSIM gates Eq.~(\ref{fsim}) for different layers of entanglers $\hat{U}_{E,1}$ and $\hat{U}_{E,2}$, i.e., all the FSIM gates in the PQC shown in Fig.~\ref{fig1}(b) are the same. For the layers of single-qubit rotations, we adopt the protocol where the even (odd) qubits have the same angle $\overrightarrow{\alpha}^{(P)}$ ($\overrightarrow{\beta}^{(P)}$). The reason for choosing this protocol is analyzed in Appendix B from the perspective of the expressibility~\cite{expr_add} of PQCs and the spatial symmetry.

Then, the final state after the $p$-layer PQC based on the ALA on the initial state $|\psi_{0}\rangle = |00...0\rangle$ is written as
\begin{eqnarray}
|\psi(\vec{x})\rangle = \prod_{l=1}^{p}[\hat{U}_{s}^{(2l)}\hat{U}_{E,2}\hat{U}_{s}^{(2l-1)}\hat{U}_{E,1}]\hat{U}_{s}^{(0)}|\psi_{0}\rangle
\label{final_state}
\end{eqnarray}
with 
\begin{eqnarray}
\hat{U}_{s}^{(k)}\equiv \prod_{j\in\text{odd}} \hat{U}_{j}(\vec{\alpha}^{(k)}) \prod_{j\in\text{even}}\hat{U}_{j}(\vec{\beta}^{(k)}) 
\label{us}
\end{eqnarray}
being a layer of single-qubit rotations (\ref{single_1}). Here, 
\begin{eqnarray}
\hat{U}_{s}^{(0)} = \prod_{j\in\text{odd}}e^{-i\hat{\sigma}_{j}^{z}\alpha_{1}^{(0)}}e^{-i\hat{\sigma}_{j}^{x}\alpha_{2}^{(0)}}\prod_{j\in\text{even}}e^{-i\hat{\sigma}_{j}^{z}\beta_{1}^{(0)}}e^{-i\hat{\sigma}_{j}^{x}\beta_{2}^{(0)}}  
\label{us0}
\end{eqnarray}
and the initial single-qubit rotation without the first $Z$ rotation in Eq.~(\ref{single_1}), due to the chosen initial state $|\psi_{0}\rangle$. In the PQC with $p$ layers, in addition to four variational parameters in the initialization, there are $12p$ variational parameters for the single-qubit rotations, and only two variational parameters for the entanglers.

After introducing the PQC, we can implement the optimization algorithms, such as the BFGS method~\cite{vq_5}, to search for the optimum parameters in the PQC for the state with the lowest squeezing parameter. We present details of the optimization by using the BFGS method in Appendix C. Note that the squeezing can be visualized via the Husimi $Q$ function $Q(\Theta,\Phi)\propto |\langle\psi(\vec{x})|\Theta,\Phi\rangle|^{2}$ with $|\Theta,\Phi\rangle\equiv \otimes_{j=1}^{N}(\cos\frac{\Theta}{2}|0\rangle_{j} + \sin\frac{\Theta}{2}e^{-i\Phi}|1\rangle_{j})$. Figure~\ref{fig1}(c) plots the $Q$ function of the optimized state obtained from the variational spin-squeezing algorithm workflow with a depth of PQC $p=3$ and a qubit number $N=12$. The squeezing parameter of the optimized spin-squeezed state is $\xi_{\text{opt}}^{2}\simeq 0.2$, which is comparable to the lowest value of the squeezing parameter for TAT, i.e., $\xi_{\text{TAT}}^{2}\simeq 0.24$ (see Appendix D for the definitions of OAT and TAT and related numerical results).

We numerically simulate the variational spin-squeezing algorithm workflow with the PQC based on the ALA, using the BFGS as the optimization algorithm for total $12p + 6$ variational parameters, and study the optimized linear Ramsey squeezing parameter $\xi^{2}_{L,\text{opt}}$ as a function of the depth of the PQC $p$. Here, we adopt the PQC with periodic boundary conditions (PBCs), as displayed in Fig.~\ref{fig1}(a) and (b). 

Figure~\ref{fig2}(a) shows the results with the system size up to $N=20$. It is shown that for $p=1$ and $2$, the results for different $N$ give the same value of $\xi^{2}_{L,\text{opt}}$. However, with increasing $p$, 
the values of $\xi^{2}_{L,\text{opt}}$ diverge for different $N$. Actually, $\xi^{2}_{L,\text{opt}}$ will approach the fundamental limit of 
the linear Ramsey squeezing parameter $\xi^{2}_{\text{lim}} = 2/(N+2)$ [see the inset of Fig.~\ref{fig2}(a) for the results of $\xi^{2}_{L,\text{opt}}$ with a deeper $p$ achieving $\xi^{2}_{\text{lim}}$]. 
Figure~\ref{fig2}(b) displays $\xi^{2}_{L,\text{opt}}/\xi^{2}_{L,\text{TAT}}$ as a function of $p$. We can define the $p^{*}$ as the most shallow depth achieving $\xi^{2}_{L,\text{opt}}/\xi^{2}_{L,\text{TAT}}<1$. We plot $p^{*}$ as a function of $N$ in the inset of Fig.~\ref{fig2}(b), showing a slow (approximately linearly) growth of $p^{*}$ as $N$ increases. 

Note that OAT or TAT, with infinite-range interactions (see Appendix D), can also be realized by directly decomposing the infinite-range circuit into nearest-neighbor ones, which is known as the \emph{line swap strategy}~\cite{Weidenfeller2022scalingofquantum,Harrigan:2021we}. By using this strategy, a $N$-qubit infinite-range circuit can be decomposed in $N$-layers of nearest-neighbor two-qubit gates. Although the depth of the circuit based on the line swap strategy also linearly scales with the qubit number $N$, to achieve OAT or TAT via the variational spin-squeezing algorithms, the required depth of the PQC is shallower [see the inset of Fig.~\ref{fig2}(b)].

We then discuss whether a highly spin-squeezed state can be generated via the variational algorithm without the PBC of the PQC. We first consider a simple case where the PQC is the same as Fig.~\ref{fig1}(b) but with open boundary conditions (OBCs), i.e., the two-qubit gate $\text{FSIM}(\theta,\phi)_{1,N}$ is absent. The results are
displayed in Fig.~\ref{fig2}(c), showing that the $\xi^{2}_{L,\text{opt}}$ obtained from the PQC with OBC is worse than that for the PQC with PBC. 

Next, we consider a protocol, where the single-qubit rotations depends on the index of qubits and the layers (see Appendix B for a schematic representation of the PQC with site-dependent rotations). As shown in Fig.~\ref{fig2}(c), with the OBC, although one can obtain a lower value of $\xi^{2}_{L,\text{opt}}$ via the protocol with site-dependent rotations, the variational algorithm with an open-boundary PQC cannot generate an optimized spin-squeezed state with a squeezing parameter comparable with the case of PBC, suggesting that the PBC plays a key role in designing the PQCs for efficient variational spin-squeezing algorithms. This can be interpreted by the fact that both the OAT and TAT Hamiltonian, i.e., $\hat{H}_{\text{OAT}}\propto \hat{J}^{2}_{z}$ and $\hat{H}_{\text{TAT}}\propto \hat{J}^{2}_{x} - \hat{J}^{2}_{y}$, have the PBC with cyclic permutation symmetry. Thus, it is expected that the PQCs for efficient variational spin-squeezing algorithms should fulfill the symmetry.

In addition, we consider the PQC, where the angles $(\theta,\phi)$ of the FSIM gates (\ref{fsim}) in each layer of entanglers are different, i.e., the protocol with layer-dependent entanglers. We present the schematic of the PQC in Appendix B. As displayed in Fig.~\ref{fig2}(c), with PBC, the optimized squeezing parameters $\xi^{2}_{L,\text{opt}}$ for the protocol with layer-dependent entanglers are only sightly smaller than those for the conventional protocol. However, the protocol with layer-dependent entanglers requires more experimental cost than the protocol shown in Fig.~\ref{fig1}(b) since the calibration of two-qubit FSIM gates with different angles $(\theta,\phi)$ 
is more demanding than single-qubit gates~\cite{PhysRevLett.123.210501}. Consequently, the PQC shown in Fig.~\ref{fig1}(b) is more experimentally feasible. In Appendix E, in order to guide subsequent experimental investigations, we present the optimized angles $(\theta_{\text{opt}}, \phi_{\text{opt}})$ of the FSIM gates (\ref{fsim}) for the variational spin-squeezing algorithms with the PQC shown in Fig.~\ref{fig1}(b).

The linear squeezing parameter can be generalized to the second-order non-linear squeezing parameter by adopting $\hat{\mathbf{S}} = \hat{\mathbf{S}}_{NL} = (\hat{J}_{x},\hat{J}_{y},\hat{J}_{z},\hat{J}_{x}^{2},\hat{J}_{y}^{2},\hat{J}_{z}^{2},\hat{J}_{xy}^{2},\hat{J}_{yz}^{2},\hat{J}_{zx}^{2})$ in Eq.~(\ref{ssp}), which can characterize the entanglement of non-Gaussian states~\cite{PhysRevLett.122.090503}. Here we consider the non-linear squeezing parameter $\xi^{2}_{NL}$ as the objective function and employed the variational spin-squeezing algorithm workflow with the PQC in Fig.~\ref{fig1}(b) to obtain the optimized non-linear squeezing parameter $\xi^{2}_{NL, \text{opt}}$. 

As displayed in Fig.~\ref{fig2}(d), for $N=10$, $\xi^{2}_{NL, \text{opt}}$ can achieve the lowest value of $\xi^{2}_{NL}$ obtained from OAT and TAT with the depths $p=4$ and $7$, respectively. In comparison with the linear squeezing parameter, a deeper depth of the PQC is required for achieving the lowest $\xi^{2}_{NL}$ of TAT. It has been verified that there is a hierarchy 
\begin{eqnarray}
\xi^{-2}_{L}\leq\xi^{-2}_{NL}\leq F/N
\label{hierarchy}
\end{eqnarray}
with $F$ being the quantum Fisher information~\cite{PhysRevLett.122.090503}. Consequently, the inverse of the second-order squeezing parameter $\xi^{-2}_{NL}$ can estimate the \emph{lower bound} of $F/N$. For $p=7$, $\xi^{-2}_{NL, \text{opt}}\simeq 6.87$. Because the violation of the inequality $F/N\leq \kappa$ signals $(\kappa+1)$-partite entanglement~\cite{RevModPhys.90.035005}, the optimized state with $p=7$ \emph{at least} has $7$-partite entanglement. 

\begin{figure}
	\centering
	\includegraphics[width=1\linewidth]{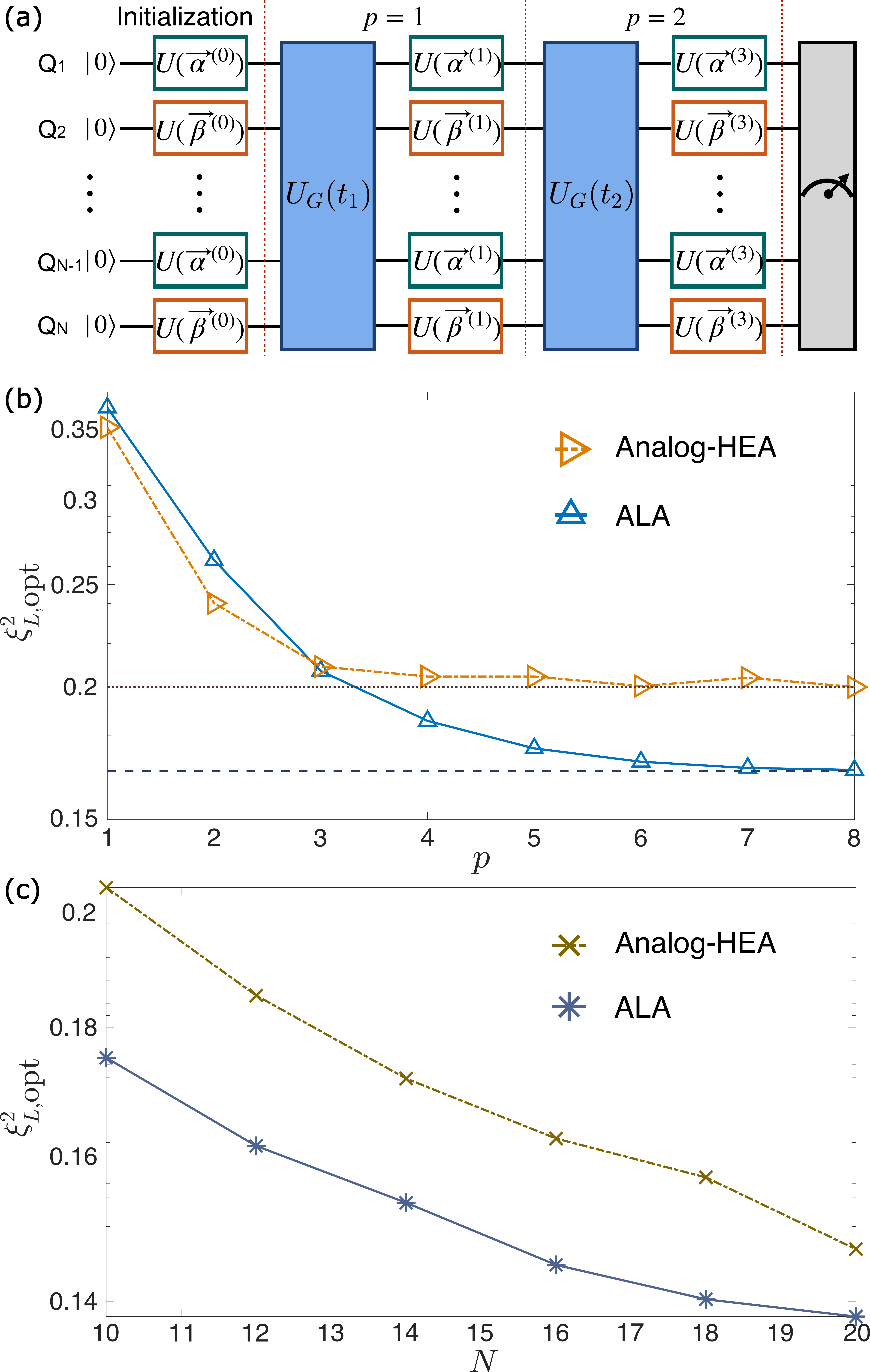}\\
	\caption{ (a) Schematic of the PQC for the analog-HEA protocol with the number of layers $p=2$. (b) Semilogarithmic plot of the optimized linear Ramsey squeezing parameter $\xi^{2}_{L,\text{opt}}$ versus the depth $p$ of the PQC for $N=10$. The dotted horizontal line is the saturated value of $\xi^{2}_{L,\text{opt}}$ for the analog-HEA protocol, i.e., $\xi^{2}_{L,\text{opt}} \simeq 0.2$. The dashed horizontal line represents the fundamental limit of the squeezing parameter with $N=10$, i.e., $\xi^{2}_{\text{lim}}\simeq 0.167$. (c) The $\xi^{2}_{L,\text{opt}}$ as a function of the number of qubits $N$ with the depth of PQCs $p=5$. Here, ALA refers to alternating layered ansatz. }\label{fig3}
\end{figure}

\begin{figure}[]
  \centering
  \includegraphics[width=1\linewidth]{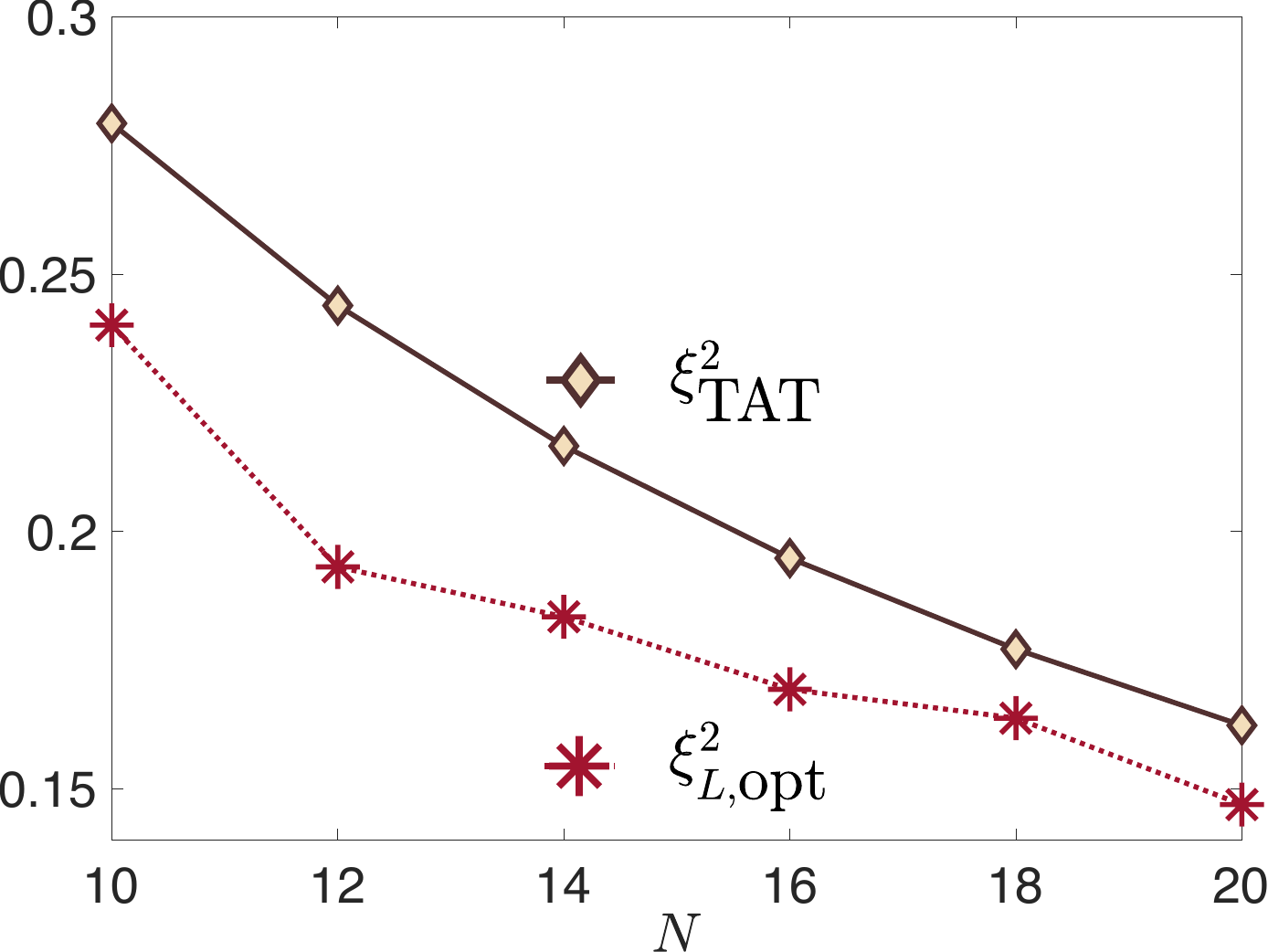}\\
  \caption{The optimized linear Ramsey squeezing parameter $\xi^{2}_{L,\textrm{opt}}$ obtained from the variational spin-squeezing algorithm based on the PQC shown in Fig.~\ref{fig3}(a) with depth $p = p^{*}$, for different numbers $N$ of qubits, in comparison with the $\xi^{2}_{\text{TAT}}$. }\label{s_3_2}
\end{figure}

\begin{figure*}[]
  \centering
  \includegraphics[width=1\linewidth]{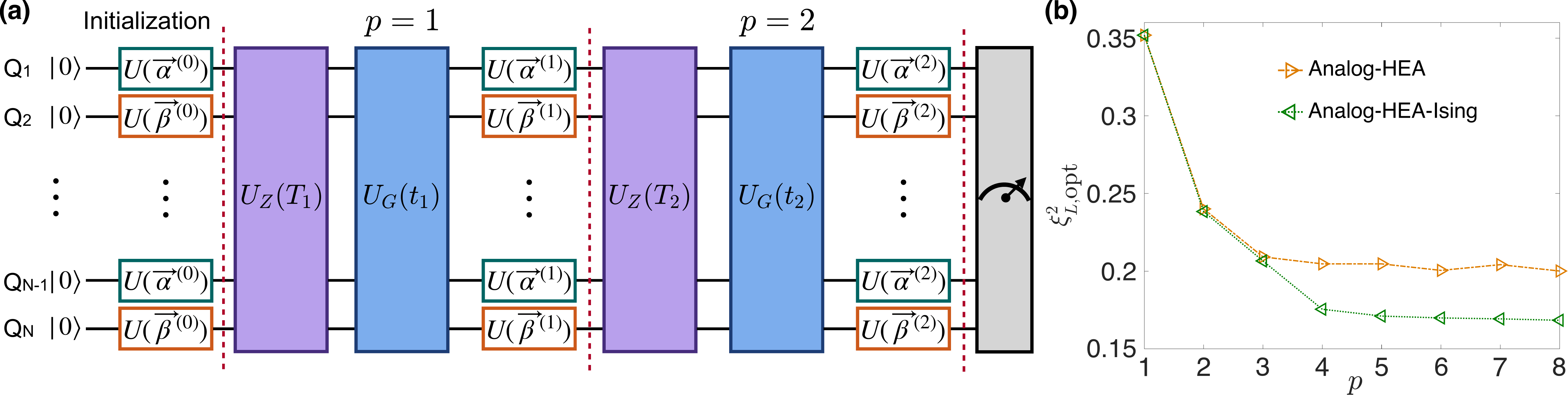}\\
  \caption{(a) Schematic of the PQC for the analog-HEA protocol consisting of both the time evolution of the XY model $\hat{U}_{G}(t_{j})$ and the Ising model $\hat{U}_{Z}(T_{j})$ (the analog-HEA-Ising protocol). Here, the depth of the PQC is $p=2$. (b) The optimized linear Ramsey squeezing parameter $\xi^{2}_{L,\text{opt}}$ versus the depth $p$ of the PQC for $N=10$. Here, both the analog-HEA-Ising protocol and the analog-HEA protocol are considered.  }\label{re_1}
\end{figure*}

\section{Variational spin-squeezing algorithm with analog quantum circuits}

For the PQC as shown in Fig.~\ref{fig1}(b), between two layers of single-qubit rotations, the entanglers $\hat{U}_{E,1}$ and $\hat{U}_{E,2}$ only make local qubit pairs entangled, as a typical digital quantum circuit. In contrast, for the hardware-efficient ansatz (HEA) introduced in Ref.~\cite{sc8}, all qubits are entangled between two layers of single-qubit rotations, as shown in Fig.~\ref{fig3}(a). Here, instead of using a gate-model-based way to construct the global entangler, i.e., $\hat{U}_{G} = \hat{U}_{E,1} \hat{U}_{E,2}$~\cite{sc8}, we focus on the analog quantum process, where the global entangler is realized by letting all qubits evolve under a tailored Hamiltonian $\hat{H}_{T}$.

According to the results in Fig.~\ref{fig2}(c), a PQC with PBC and layer-dependent entanglers has a better performance. Thus, for the design of analog quantum circuits based on the HEA, i.e., the analog-HEA protocol, we still adopt PBC and the layer-dependent global entangler $\hat{U}_{G}(t_{j})=\exp(-i\hat{H}_{T}t_{j})$ with variational parameters $t_{j}$ ($j=1,2,...,p$ for a $p$-layer PQC). We consider 1D analog superconducting circuits with nearest-neighbor interactions~\cite{Kockum2019,sc5,sc6,sc9,sc10,Kannan:2020uv,PhysRevB.75.140515,Gu:2017ut} as an example, and the tailored Hamiltonian can be written as (see Appendix F for details of the experimental realization)
\begin{equation}
\hat{H}_{T} = \sum_{i=1}^{N}(\hat{\sigma}_{i}^{x}\hat{\sigma}_{i+1}^{x} + \hat{\sigma}_{i}^{y}\hat{\sigma}_{i+1}^{y} ),\label{HXY} 
\end{equation}
which is known as the 1D XY model. Overall, the final parameterized state obtained from the $p$-layer PQC of the analog-HEA protocol is
\begin{eqnarray}
|\psi(\vec{x})\rangle = \prod_{l=1}^{p}[\hat{U}_{s}^{(l)}\hat{U}_{G}(t_{l})]\hat{U}_{s}^{(0)}|\psi_{0}\rangle, 
\label{final_state_add}
\end{eqnarray}
where the initial state $|\psi_{0}\rangle$ and the single-qubit rotations  $\hat{U}_{s}^{(l)}$ and $\hat{U}_{s}^{(0)}$ are the same as those in Eq.~(\ref{final_state}). More precisely, the PQC  $\hat{U}(\vec{x})=\prod_{l=1}^{p}[\hat{U}_{s}^{(l)}\hat{U}_{G}(t_{l})]\hat{U}_{s}^{(0)}$ is an analog-digital circuit, because the layers of single-qubit rotations $\hat{U}_{s}^{(l)}$ can be regarded as digital blocks~\cite{PhysRevA.101.022305}.

With the linear Ramsey squeezing parameter $\xi^{2}_{L}$ being the objective function, after the optimization, we can obtain the lowest squeezing parameter $\xi^{2}_{L,\text{opt}}$ for the analog-HEA protocol. As shown in Fig.~\ref{fig3}(b), when increasing the depth $p$, the $\xi^{2}_{L,\text{opt}}$ obtained from the ALA tends to the fundamental limit $\xi^{2}_{\text{lim}}$, while $\xi^{2}_{L,\text{opt}}$ obtained from the analog-HEA protocol saturates to a larger value, indicating that the variational spin-squeezing algorithm based on the ALA outperforms with the analog-HEA protocol.  

As shown in the inset of Fig.~\ref{fig2}(b), TAT can be achieved by using the variational spin-squeezing algorithm based on the ALA with the depth $p^{*}=2$, $3$, $3$, $4$, $4$, and $5$ for $N=10$, $12$, $14$, $16$, $18$, and $20$. Here, in Fig.~\ref{s_3_2}, we plot the $\xi^{2}_{L,\textrm{opt}}$ obtained from the analog-HEA protocol with $p=p^{*}$ for different $N$. It is seen that the shallow-depth analog-HEA protocol can still generate a spin-squeezed state better than that obtained from the TAT, which provides the possibility for generating strong squeezing on near-term analog quantum computers. 

\begin{figure*}
	\centering
	\includegraphics[width=1\linewidth]{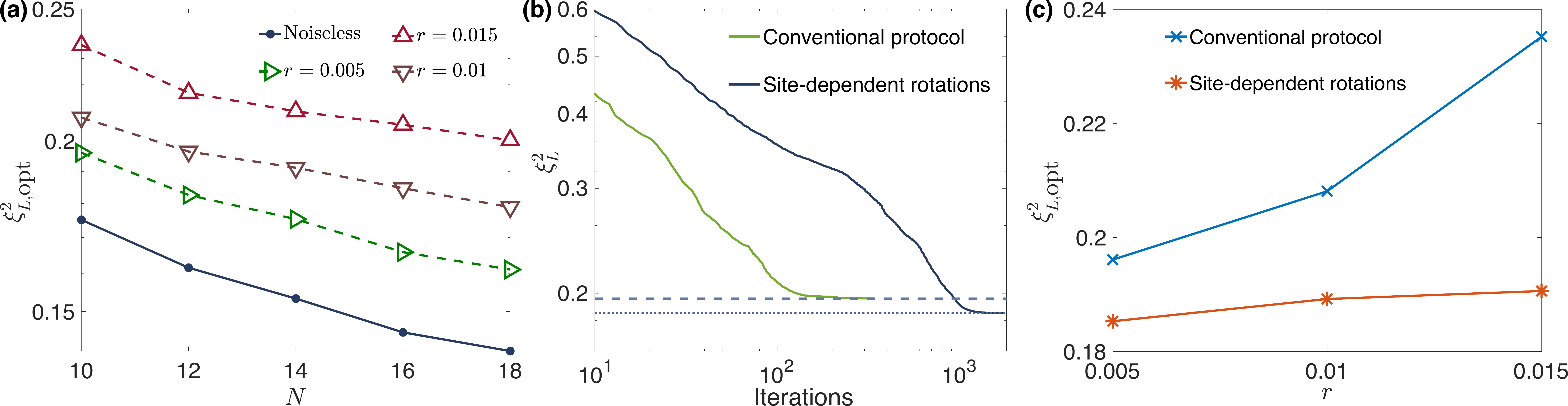}\\
	\caption{ (a) The optimized linear Ramsey squeezing parameter $\xi^{2}_{L,\text{opt}}$ obtained from the variational spin-squeezing algorithm based on the ALA as a function of the number $N$ of qubits with depth $p=5$ and different coherent errors $r$. (b) For $N=10$ and $r=0.005$, the squeezing parameter $\xi^{2}_{L}$ versus iterations during the optimization, for both the conventional ALA protocol and the ALA with site-dependent rotations. (c) For $N=10$ and $p=5$, the optimized squeezing parameter $\xi^{2}_{L,\text{opt}}$ with different coherent errors $r$. Here, we consider both the conventional ALA protocol and the ALA with site-dependent rotations. Note that (a) has a vertical logarithmic axis, and (b) has both logarithmic axes.}\label{fig4}
\end{figure*}


The aforementioned analog-HEA protocol only considers the unitary evolution under the 1D XY model (\ref{HXY}), because it can be directly realized in analog superconducting circuits~\cite{Kockum2019,sc5,sc6,sc9,sc10,Kannan:2020uv,PhysRevB.75.140515,Gu:2017ut} .  For the ALA protocol with the PQC shown in Fig.~\ref{fig1}(b), the entanglers consist of the FSIM gates (\ref{fsim}), which can be rewritten as 
\begin{eqnarray}
\text{FSIM}(\theta,\phi) = \exp[-i\frac{\theta}{2} (\hat{\sigma}^{x}\otimes\hat{\sigma}^{x} + \hat{\sigma}^{y}\otimes\hat{\sigma}^{y})] \\ \nonumber
\times \exp[-i\frac{\phi}{4}(\hat{I} - \hat{\sigma}^{z})\otimes(\hat{I} - \hat{\sigma}^{z})]
\label{fsim_re}
\end{eqnarray}
($\hat{I}$ is the two-dimensional identity matrix). It is seen that both the XY-type interactions $\exp[-i\frac{\theta}{2} (\hat{\sigma}^{x}\otimes\hat{\sigma}^{x} + \hat{\sigma}^{y}\otimes\hat{\sigma}^{y})]$ and Ising interactions $\exp(-i\frac{\phi}{4} \hat{\sigma}^{z}\otimes\hat{\sigma}^{z})$ exist in the FSIM gates. However, for the  analog-HEA protocol with the PQC in Fig.~\ref{fig3}(a), the Ising interactions are absent.   

Next, we explore the variational spin-squeezing algorithm based on the PQC shown in Fig.~\ref{re_1}(a), with an additional entangler 
\begin{eqnarray}
\hat{U}_{z}(T_{j}) = \exp[-iT_{j}(\sum_{k=1}^{N} \hat{\sigma}^{z}_{k}\hat{\sigma}^{z}_{k+1})]
\label{uzz}
\end{eqnarray}
as the global Ising interaction, i.e., the analog-HEA-Ising protocol. The final parameterized state obtained from the $p$-layer PQC shown in Fig.~\ref{re_1}(a) is 
\begin{eqnarray}
|\psi(\vec{x})\rangle = \Pi_{l=1}^{p} [\hat{U}_{s}^{(l)}\hat{U}_{G}(t_{l})\hat{U}_{z}(T_{l})] \hat{U}_{s}^{(0)}|\psi_{0}\rangle,
\label{psi_x_add}
\end{eqnarray}
where $|\psi_{0}\rangle$, $\hat{U}_{s}^{(0)}$, and $\hat{U}_{s}^{(l)}$ are the same as those in Eq.~(\ref{final_state_add}).  

In Fig.~\ref{re_1}(b), we display the optimized squeezing parameter $\xi^{2}_{L,\text{opt}}$ for the analog-HEA-Ising protocol, in comparison with the analog-HEA protocol shown in Fig.~\ref{fig3}(a).  It is seen that the $\xi^{2}_{L,\text{opt}}$ obtained from the analog-HEA-Ising protocol can achieve lower values, which tend to the fundamental limit $\xi^{2}_{\text{lim}}\simeq 0.167$ with increasing $p$. The results shown in Fig.~\ref{re_1}(b) indicate that the Ising interactions are indispensable for the design of variational spin-squeezing algorithms, which can achieve the fundamental limit of the linear Ramsey squeezing parameter.

\section{Experimental imperfections}

We now analyze the influence of experimental imperfections on variational spin-squeezing algorithms. We focus on the algorithm based on the ALA, which is shown in Fig.~\ref{fig1}(b), since it outperforms the analog-HEA protocol with $p=5$ of the PQC [see Fig.~\ref{fig3}(c)]. Here we study the influence of the coherent errors in the FSIM gate (\ref{fsim}). In experiments, the coherent errors can be induced by the uncertainty of $\theta$ and $\phi$ in Eq.~(\ref{fsim}), and the additional phases $\Delta_{+}$, $\Delta_{-}$, and $\Delta_{-,\text{off}}$~\cite{coherent_error_OTOC}. One can see Appendix G for the formulation of the FSIM gate with coherent errors, i.e., $\text{FSIM}_{\text{exp.}}(\theta_{*},\phi_{*},\Delta_{+}, \Delta_{-}, \Delta_{-,\text{off}})$. The coherent error can be quantified by
\begin{eqnarray}
r = 1-\frac{1}{16}|\text{Tr}(\text{FSIM}_{\text{exp.}}^{\dagger}\cdot\text{FSIM}_{\text{target}})|^{2}, 
\label{error}
\end{eqnarray}
where the target FSIM gate is $\text{FSIM}_{\text{target}} = \text{FSIM}(\theta_{\text{opt}},\phi_{\text{opt}})$ with the optimized parameters $(\theta_{\text{opt}},\phi_{\text{opt}})$ obtained from the variational algorithm without the coherent error, and $\text{FSIM}_{\text{exp.}}$ represents the FSIM gate with the coherent error.


As shown in Fig.~\ref{fig4}(a), with coherent errors, the squeezing generated by the variational algorithm becomes weaker. 
When the coherent errors in the entanglers are considered, the $\text{FSIM}$ gates of different qubit pairs are not identical (see Appendix G), and the partial permutation symmetry in the PQC as shown in Fig.~\ref{fig1}(b), where all the $\text{FSIM}$ gates are the same, is absent for the PQC with coherent errors. Consequently, according to the results of the PQC with site-dependent rotations [see Fig.~\ref{fig2}(c)], one can expect a lower $\xi^{2}_{L,\text{opt}}$ by employing the PQC with site-dependent rotations than its conventional version. As shown in Fig.~\ref{fig4}(b), the optimization trajectory for the saturation of $\xi^{2}_{L}$ to its optimized value $\xi^{2}_{L,\text{opt}}$ is longer for the protocol with site-dependent rotations compared to the conventional protocol, due to its larger number of variational parameters. Moreover, the influence of the coherent errors on the generated squeezing is suppressed by using the protocol with site-dependent rotations [see Fig.~\ref{fig4}(c)].


\section{DISCUSSIONS}

We have developed variational spin-squeezing algorithms for 1D quantum devices with nearest-neighbor interactions. We designed the PQC enlightened by the ALA, suitable for digital quantum computers, as well as the PQC based on the HEA, which can be naturally realized in analog superconducting circuits. We 
demonstrated that variational spin-squeezing algorithms with both digital and analog versions of the PQC can generate the spin-squeezed states comparable to the best spin-squeezed state obtained from TAT. Our work sheds light on the variational optimization of quantum sensors~\cite{vqe_sss5,vqe_sss6,RevModPhys.89.035002} for 1D quantum devices with nearest-neighbor interactions. 

The variational spin-squeezing algorithms proposed in this work pave the way for experimentally achieving the strongest squeezing generated from TAT.
The efficient numerical simulation of the PQC shown in Fig.~\ref{fig1}(b) with PBC for large system sizes is an intractable problem, because the matrix-product-state-based algorithm is less efficient for PBC~\cite{RevModPhys.77.259}. Consequently, testing variational spin-squeezing algorithms on actual experimental platforms is desirable, because it can extend the results shown in the inset of Fig.~\ref{fig2}(b) to larger system sizes and demonstrate the scalability of variational spin-squeezing algorithms. 

For the analog version of the variational spin-squeezing algorithm, the global entangler is implemented by the unitary dynamics under the tailored Hamiltonian that describes an array of superconducting qubits with nearest-neighbor capacitive couplings. A further direction is to study the performance of the variational spin-squeezing algorithm based on other experimental platforms such as the Rydberg-dressed atoms~\cite{Henriet2020quantumcomputing,Browaeys:2020tl,PhysRevX.7.041063,PhysRevLett.124.063601} and cavity quantum electrodynamics systems~\cite{PhysRevX.2.021007,Qin4868,PhysRevA.101.053818}. 



\begin{acknowledgments}
We acknowledge the discussions with Markus Heyl and Xiao Yuan. This work was supported by National Natural Science Foundation of China (Grants Nos. 92265207, T2121001, 11934018), Innovation Program for Quantum Science and Technology  (Grant No. 2-6), Beijing Natural Science Foundation (Grant No. Z200009), and Scientific Instrument Developing Project of Chinese Academy of Sciences (Grant No. YJKYYQ20200041), Nippon Telegraph and Telephone Corporation (NTT) Research,
the Japan Science and Technology Agency (JST) [via the Quantum Leap Flagship Program (Q-LEAP), and the Moonshot R\&D  Grant Number JPMJMS2061], the Asian Office of Aerospace Research and Development (AOARD) (via Grant No. FA2386-20-1-4069), and
the the office of Naval Research (ONR).

\end{acknowledgments}

\appendix



\section{Measurement of spin squeezing parameters}

 \begin{figure}[]
  \centering
  \includegraphics[width=1\linewidth]{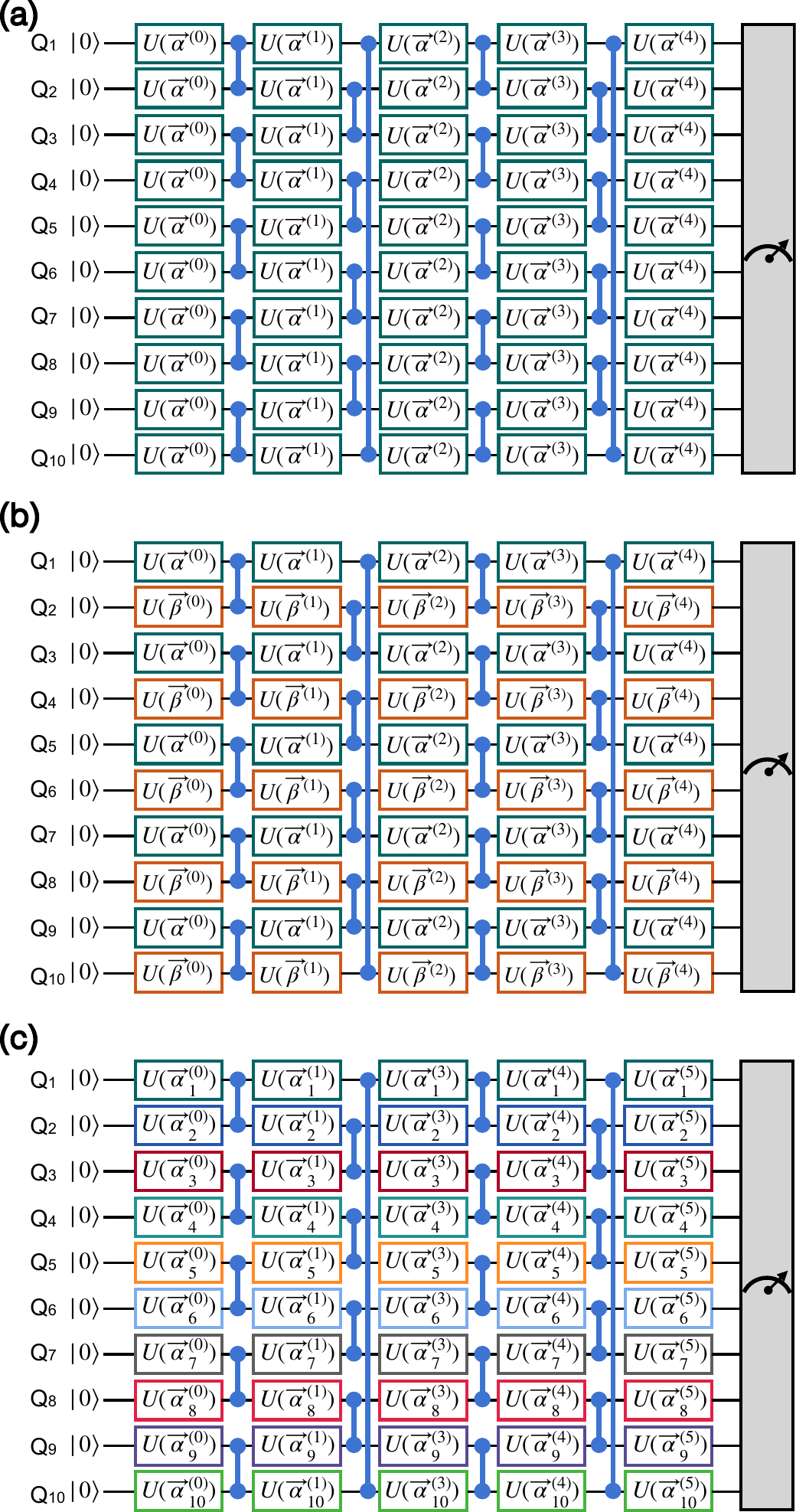}\\
  \caption{(a) A PQC with global single-qubit gates and PBC. (b) A PQC which is similar to that shown in the Fig~\ref{fig1}(b) but with the system size $N=10$.   (c) A PQC with site-dependent single-qubit gates and PBC.  Here, the two-qubit gate notation refers to the FSIM gate defined by Eq.~(\ref{fsim}). }\label{s_1_1}
\end{figure}

\begin{figure}[]
  \centering
  \includegraphics[width=1\linewidth]{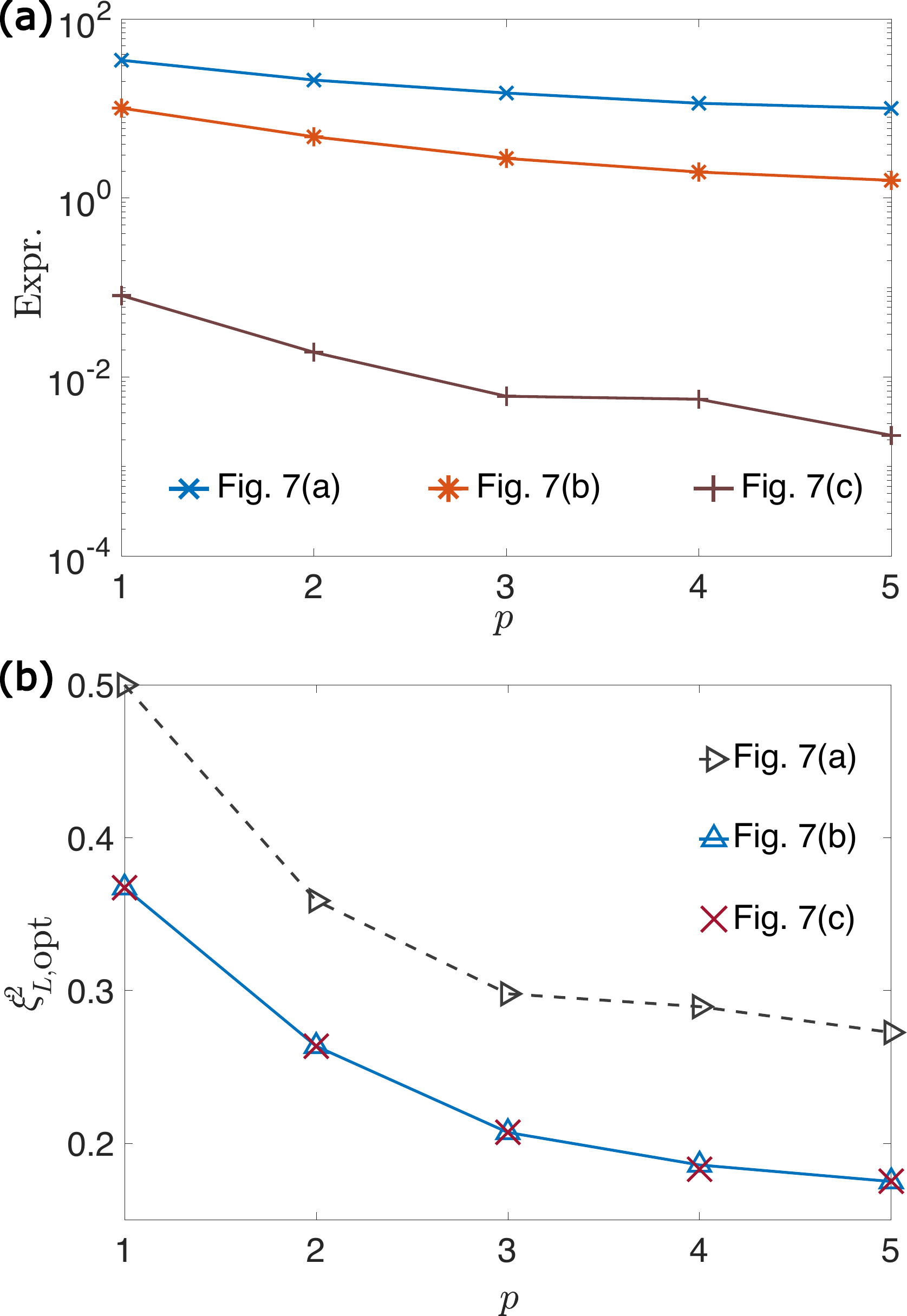}\\
  \caption{(a) For the PQCs shown in Fig.~\ref{s_1_1}, the  expressibility as a function of the depth of the PQC $p$. (b) The optimized linear Ramsey squeezing parameter $\xi^{2}_{L,\text{opt}}$, obtained from the variational spin-squeezing algorithms based on the the PQCs shown in Fig.~\ref{s_1_1}, as a function of $p$. }\label{s_1_2}
\end{figure}

\begin{figure}[]
  \centering
  \includegraphics[width=1\linewidth]{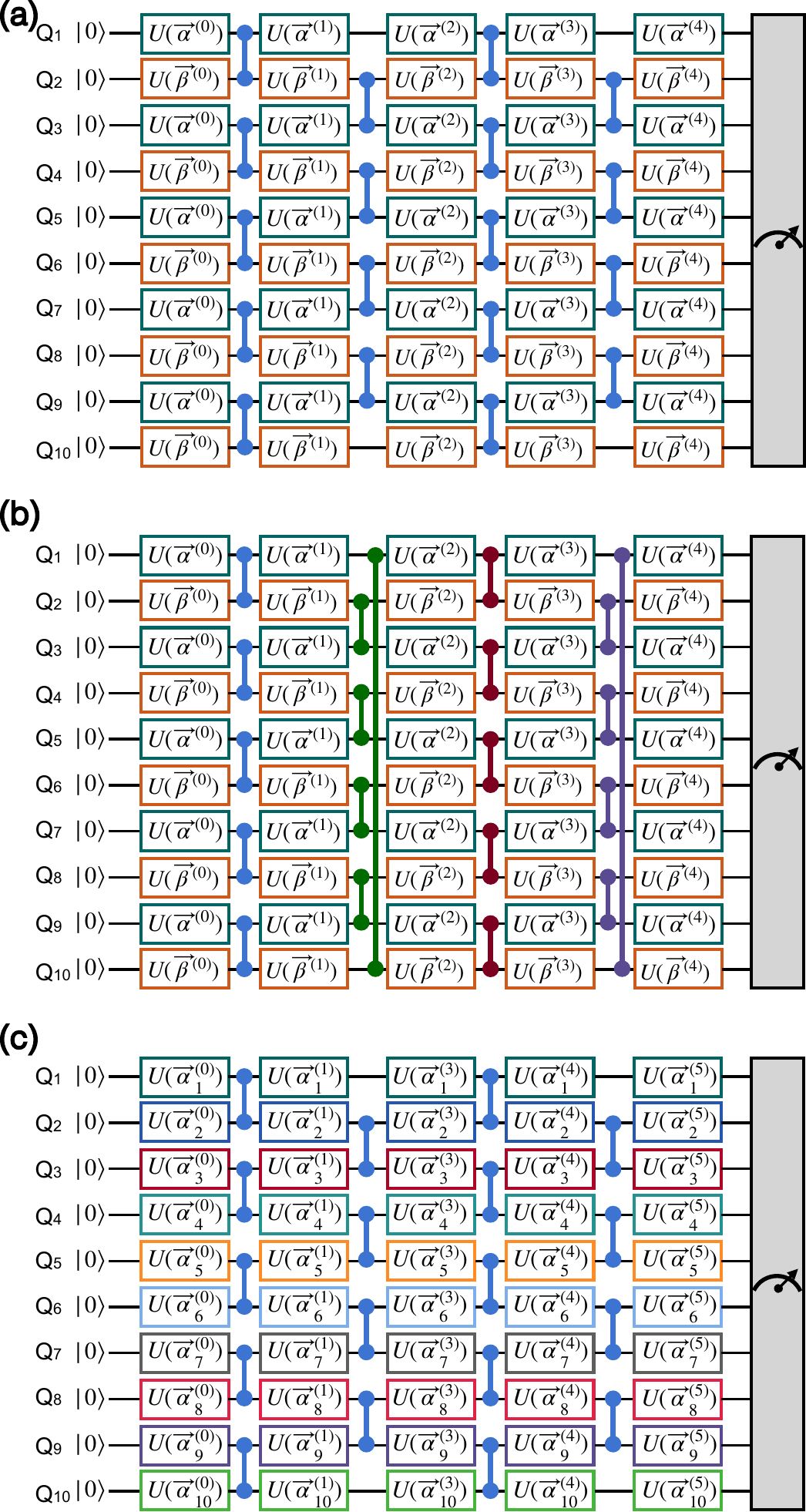}\\
  \caption{(a)  A PQC which is similar to that shown in the Fig~\ref{s_1_1}(b), but with the open boundary condition (OBC).  (b)  A PQC which is similar to that shown in the Fig~\ref{s_1_1}(b), but with layer-dependent entanglers. (c)  A PQC with site-dependent single-qubit gates and OBC. Here, the two-qubit gate notation refers to the FSIM gate defined by Eq.~(\ref{fsim}). }\label{s_1_3}
\end{figure}

We first present the measurement scheme of the linear spin squeezing parameter $\xi_{L}^{2}$ with a set of collective spin operators $\hat{\mathbf{S}}_{L} = (\hat{J}_{x},\hat{J}_{y},\hat{J}_{z})$. According to Eq.~(\ref{ssp}), the parameter $\xi_{L}^{2}$ can be calculated with the matrix 
\begin{eqnarray}
\mathbb{C}[\hat{\mathbf{S}}_{L}] = 
\begin{pmatrix}
0 & \langle\hat{J}_{z}\rangle & -\langle\hat{J}_{y}\rangle \\
-\langle\hat{J}_{z}\rangle & 0 & \langle\hat{J}_{x}\rangle \\
\langle\hat{J}_{y}\rangle & -\langle\hat{J}_{x}\rangle & 0
\end{pmatrix},  
\label{c_more}
\end{eqnarray}
and 
\begin{eqnarray}
\mathbb{V}[\hat{\mathbf{S}}_{L}] = 
\begin{pmatrix}
(\Delta \hat{J}_{x})^{2} & \text{cov}(\hat{J}_{x}, \hat{J}_{y}) & \text{cov}(\hat{J}_{x}, \hat{J}_{z}) \\
\text{cov}(\hat{J}_{x}, \hat{J}_{y})  & (\Delta \hat{J}_{y})^{2} & \text{cov}(\hat{J}_{y}, \hat{J}_{z}) \\
\text{cov}(\hat{J}_{x}, \hat{J}_{z})  & \text{cov}(\hat{J}_{y}, \hat{J}_{z})  & (\Delta \hat{J}_{z})^{2}
\end{pmatrix}. 
\label{v_more}
\end{eqnarray}
For terms $\langle\hat{J}_{\gamma}\rangle$ and $(\Delta \hat{J}_{\gamma})^{2} = \langle\hat{J}_{\gamma}^{2}\rangle - (\langle\hat{J}_{\gamma}\rangle)^{2}$ $(\gamma \in \{x,y,z\})$, the measurement requires the single-shot readout along the axis $\hat{J}_{\gamma}$. Moreover, according to 
 \begin{eqnarray}
 \text{cov}(\hat{J}_{\alpha}, \hat{J}_{\beta}) = \langle \hat{J}_{\alpha\beta}^{2} \rangle - \frac{\langle\hat{J}_{\alpha}^{2}\rangle + \langle\hat{J}_{\beta}^{2}\rangle}{2} - \langle\hat{J}_{\alpha} \rangle\langle\hat{J}_{\beta} \rangle 
 \label{cov_more}
 \end{eqnarray}
with $\alpha,\beta\in\{x,y,z\}$ ($\alpha \neq \beta$) and $\hat{J}_{\alpha\beta} = (\hat{J}_{\alpha} + \hat{J}_{\beta})\sqrt{2}$, the measurement of the covariances $\text{cov}(\hat{J}_{x}, \hat{J}_{y})$, $\text{cov}(\hat{J}_{x}, \hat{J}_{z})$ and $\text{cov}(\hat{J}_{y}, \hat{J}_{z})$ requires a single-shot readout along the additional three axes $\hat{J}_{xy}$, $\hat{J}_{xz}$ and $\hat{J}_{yz}$. In short, to obtain the linear Ramsey spin squeezing parameter $\xi_{L}^{2}$, one should perform the single-shot readout measurements on each qubit along the following axes: $\{\hat{J}_{x}, \hat{J}_{y},\hat{J}_{z},\hat{J}_{xy}, \hat{J}_{yz},\hat{J}_{zx}\}$. 

Then, we present the measurement scheme of the nonlinear spin squeezing parameter $\xi_{NL}^{2}$ with a set of collective spin operators $\hat{\mathbf{S}}_{NL} = (\hat{J}_{x},\hat{J}_{y},\hat{J}_{z},\hat{J}_{x}^{2},\hat{J}_{y}^{2},\hat{J}_{z}^{2},\hat{J}_{xy}^{2},\hat{J}_{yz}^{2},\hat{J}_{zx}^{2})$. The expression of the matrices $\mathbb{V}[\hat{\mathbf{S}}_{NL}] $ and $\mathbb{C}[\hat{\mathbf{S}}_{NL}] $ are more complex. According to the analytical results in Ref.~\cite{PhysRevLett.128.150501}, to measure the parameter $\xi_{NL}^{2}$, one requires the single-shot readout measurements of the following operators 
 \begin{eqnarray}
 \hat{J}_{x}, \hat{J}_{y},\hat{J}_{z},\hat{J}_{xy}, \hat{J}_{yz},\hat{J}_{zx},  \hat{J}_{x\overline{y}}, \hat{J}_{y\overline{z}},\hat{J}_{z\overline{x}}  
 \label{o1}
 \end{eqnarray}
with $\hat{J}_{\alpha\overline{\beta}} = (\hat{J}_{\alpha}-\hat{J}_{\beta})/\sqrt{2}$, and $\hat{J}_{xy'}, \hat{J}_{yz'},\hat{J}_{zx'}$, with $\hat{J}_{\alpha\beta'} = (\hat{J}_{\alpha}+\sqrt{3}\hat{J}_{\beta})/\sqrt{2}$, and  $\hat{J}_{x\overline{y}'}, \hat{J}_{y\overline{z}'},\hat{J}_{z\overline{x}'}$
 with $\hat{J}_{\alpha\overline{\beta}'} = (\hat{J}_{\alpha}-\sqrt{3}\hat{J}_{\beta})/\sqrt{2}$,
 and 
 \begin{eqnarray}
 \hat{J}_{xyz} = \frac{\hat{J}_{x} + \hat{J}_{y} + \hat{J}_{z}}{\sqrt{3}},  \hat{J}_{\overline{x}yz} = \frac{-\hat{J}_{x} + \hat{J}_{y} + \hat{J}_{z}}{\sqrt{3}}, 
 \label{o4}
 \end{eqnarray}
 \begin{eqnarray}
 \hat{J}_{x\overline{y}z} = \frac{\hat{J}_{x} - \hat{J}_{y} + \hat{J}_{z}}{\sqrt{3}}, \hat{J}_{xy\overline{z}} = \frac{\hat{J}_{x} + \hat{J}_{y} - \hat{J}_{z}}{\sqrt{3}}. 
 \label{o6}
 \end{eqnarray}

\section{Expressibility and performance of parameterized quantum circuits with different designs}

In this section, we study the expressibility of the parameterized quantum circuits (PQCs) employed to perform variational spin squeezing algorithms. We first introduce the method of calculating the expressibility defined in Ref.~\cite{expr_add}. The PQC can be described as $\hat{U}(\vec{x})$, where $\vec{x}$ is the variational parameter vector. The first step for calculating the expressibility involves uniformly sampling 20,000 (here for $N=10$) sets of the variational parameter. The second step requires computing the overlap $F=|\langle\psi_{0}|\hat{U}(\vec{x}_{2})\hat{U}(\vec{x}_{1})|\psi_{0}\rangle|^{2}$, with  $\vec{x}_{1}$ and $\vec{x}_{2}$ denoting two different sets of the variational parameter. The third step is obtaining the probability distribution of $F$, i.e., $P(F)$. The final step is calculating the expressibility defined as 
\begin{eqnarray}
\text{Expr.} &=& D_{\text{KL}}(P(F)||P_{\text{Haar}}(F))  \\ \nonumber
&=& \sum_{F}P(F) \ln [\frac{P(F)}{P_{\text{Haar}}(F)}], 
\label{expr_def}
\end{eqnarray}
which is actually the Kullback-Liebler divergence between the distribution $P(F)$ and the distribution corresponding to the Haar random states 
$P_{\text{Haar}}(F)=(2^{N}-1)(1-F)^{2^{N}-2}$. The expressibility can quantify the ability of a PQC to capture a rich class of quantum states. For a given PQC, a smaller value of the $\text{Expr.}$ defined by Eq.~(\ref{expr_def}) indicates a higher expressibility.

We now introduce some PQCs with periodic boundary conditions (PBCs) and different designs of single-qubit rotations. In Fig.~\ref{s_1_1}(a), we show the PQC with global single-qubit gates, i.e., all qubits have the same single-qubit rotations in each layer. The $k$-th layer of the single-qubit rotations in the PQC as shown in Fig.~\ref{s_1_1}(a) can be represented as 
\begin{eqnarray}
\hat{U}_{s}^{(k)}(\vec{\alpha}^{(k)})=\Pi_{j=1}^{N}[e^{-i\hat{\sigma}_{j}^{z}\alpha_{1}^{(k)}}e^{-i\hat{\sigma}_{j}^{x}\alpha_{2}^{(k)}}e^{-i\hat{\sigma}_{j}^{z}\alpha_{3}^{(k)}}],
\label{us_1}
\end{eqnarray}
with the site-independent angles of rotations $\vec{\alpha}^{(k)} = (\alpha_{1}^{(k)},\alpha_{2}^{(k)},\alpha_{3}^{(k)})$. 

In Fig.~\ref{s_1_1}(b), we display the PQC with the same design of Fig.~\ref{fig1}(b) in contrast to other designs of the PQCs. In Fig.~\ref{s_1_1}(c), we plot the PQC with site-dependent single-qubit gates, i.e., all qubits could have different single-qubit rotations in each layers. The $k$-th layer of the single-qubit rotations in the PQC as shown in Fig.~\ref{s_1_1}(c) can be represented as
$\hat{U}_{s}^{(k)}=\Pi_{j=1}^{N} [\hat{U}^{(k)}_{j}(\vec{\alpha}^{(k)}_{j})]$ with
\begin{eqnarray}
\hat{U}^{(k)}_{j}(\vec{\alpha}^{(k)}_{j}) = e^{-i\hat{\sigma}_{j}^{z}\alpha_{1,j}^{(k)}}e^{-i\hat{\sigma}_{j}^{x}\alpha_{2,j}^{(k)}}e^{-i\hat{\sigma}_{j}^{z}\alpha_{3,j}^{(k)}},
\label{us_2}
\end{eqnarray}
and the variational parameters $\vec{\alpha}^{(k)}_{j} = (\alpha_{1,j}^{(k)},\alpha_{2,j}^{(k)},\alpha_{3,j}^{(k)})$ being the 
angles of rotations for the $j$-th qubit. 

We then calculate the expressibility of the PQCs in Fig.~\ref{s_1_1}, and plot the results in Fig.~\ref{s_1_2}(a). 
It is seen that the expressibility of the PQC in Fig.~\ref{s_1_1}(b) is higher than that of the PQC with global single-qubit rotations [see Fig.~\ref{s_1_1}(a)], and the PQC with site-dependent single-qubit rotations [see Fig.~\ref{s_1_1}(c)] has a much higher expressibility.

In Fig.~\ref{s_1_2}(b)., we plot the optimized linear Ramsey squeezing parameter $\xi^{2}_{L,\text{opt}}$ obtained from the variational algorithms based on three different PQCs. For the variational spin-squeezing algorithm based on the PQC in Fig.~\ref{s_1_1}(b), the $\xi^{2}_{L,\text{opt}}$ is smaller than that for PQC in Fig.~\ref{s_1_1}(a), which indicates that the performance of variational spin-squeezing algorithms is related to the expressibility of the chosen PQC, and a high-expressibility PQC is required for the efficient generation of spin-squeezed states. 

Nevertheless, as shown in Fig.~\ref{s_1_2}, although the expressibility of the PQC displayed in Fig.~\ref{s_1_1}(b) is lower than of the PQC in Fig.~\ref{s_1_1}(c), the obtained values of $\xi^{2}_{L,\text{opt}}$ are more or less the same. Thus, the performance of variational spin-squeezing algorithms does not solely rely on the expressibility of the chosen PQC. 

Actually, the spatial symmetry also plays an important role in the design of PQC. The major objective of the variational spin-squeezing algorithm is generating squeezing comparable to the maximum squeezing created from the two-axis twisting (TAT). The unitary evolution of the TAT has a cyclic permutation symmetry (see the following section for more details). Meanwhile, the PQC as shown in Fig.~\ref{s_1_1}(b) can be regarded as a constrained version of the PQC in Fig.~\ref{s_1_1}(c), with partial cyclic permutation symmetry imposed. Consequently,   the PQC shown in Fig.~\ref{s_1_1}(b) and the unitary evolution of the TAT share a symmetry. The symmetry can reduce the number of parameters in the PQC, and we mainly study the variational spin-squeezing algorithm based on the form of PQC as shown in Fig.~\ref{s_1_1}(b).

In Fig.~\ref{fig2}(d), we compare the performance of variational spin-squeezing algorithms based on different PQCs. To more clearly represent the additional PQCs in Fig.~\ref{fig2}(d), we plot the schematics of the PQCs in Fig~\ref{s_1_3}. The PQC shown in Fig~\ref{s_1_3}(a) and (c) are the open-boundary-condition version of the PQC shown in Fig~\ref{s_1_1}(a) and (c), respectively. The PQC as shown in Fig~\ref{s_1_3}(b) is similar to the PQC shown in Fig~\ref{s_1_1}(b), but with layer-dependent entanglers. Specifically, the entanglers in the $p$-th layer of the PQC are $\hat{U}_{E,1}^{(p)}(\theta_{1}^{(p)},\phi_{1}^{(p)})=\otimes_{i=1}^{N/2}\text{FSIM}(\theta_{1}^{(p)},\phi_{1}^{(p)})_{2i-1,2i}$ and $\hat{U}_{E,2}^{(p)}(\theta_{2}^{(p)},\phi_{2}^{(p)})=\otimes_{i=1}^{N/2}\text{FSIM}(\theta_{2}^{(p)},\phi_{2}^{(p)})_{2i,2i+1}$, with the variational parameters $(\theta_{1}^{(p)},\phi_{1}^{(p)},\theta_{2}^{(p)},\phi_{2}^{(p)})$ dependent on the number of layers. 

\begin{figure}[t]
  \centering
  \includegraphics[width=1\linewidth]{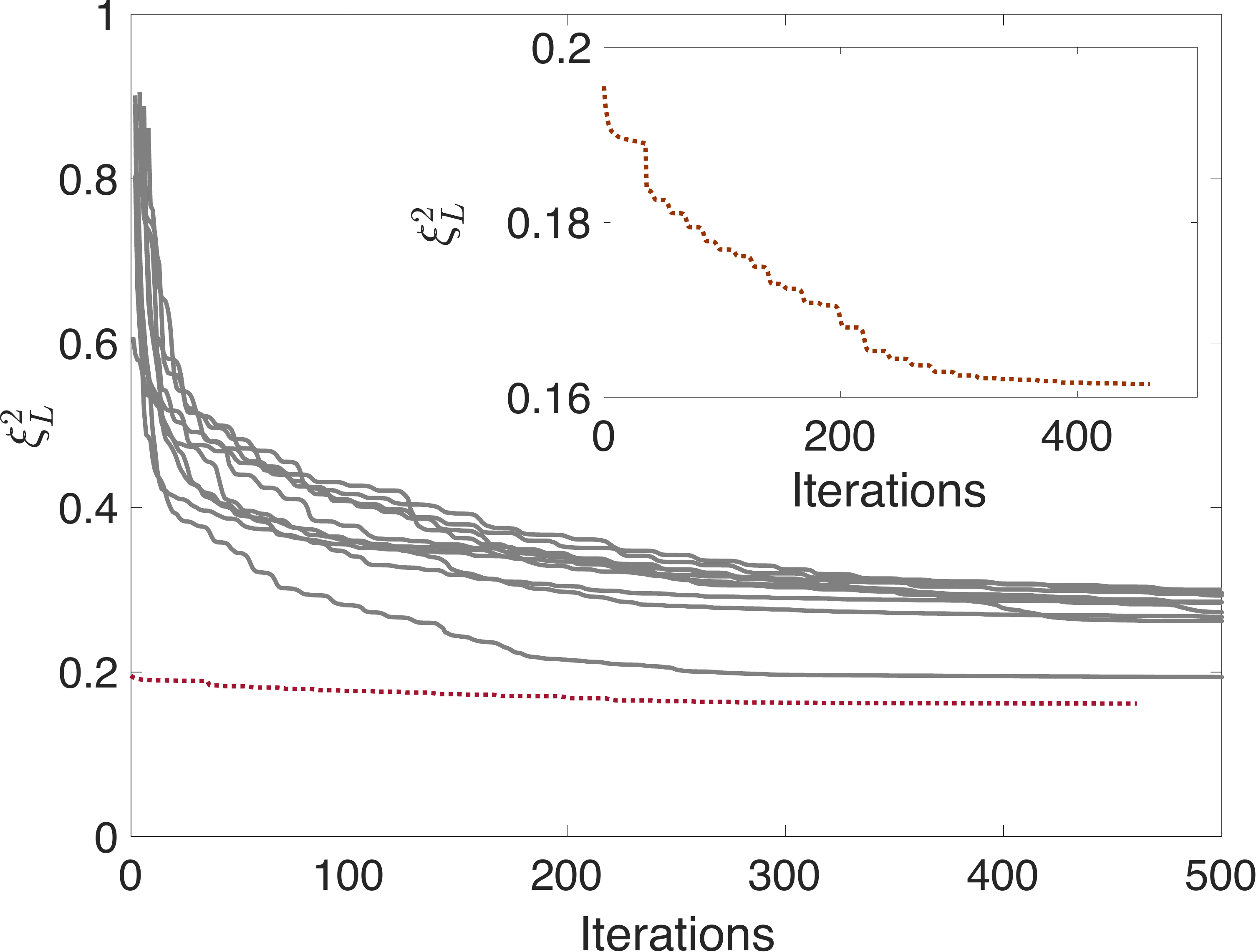}\\
  \caption{Linear Ramsey squeezing parameter $\xi_{L}^{2}$ versus iterations during the optimization. This is for the system size $N=12$ and the  $p=5$ PQC depth. The solid curves show the optimization with 10 randomly chosen initial parameters $\vec{x}_{0}^{(N=12)}$. The dotted curve shows the optimization with $\vec{x}_{0}^{(N=12)} = \vec{x}_{\text{opt}}^{(N=10)}$ (also see the inset). }\label{fig_re_2}
\end{figure}


\begin{figure}[t]
  \centering
  \includegraphics[width=1\linewidth]{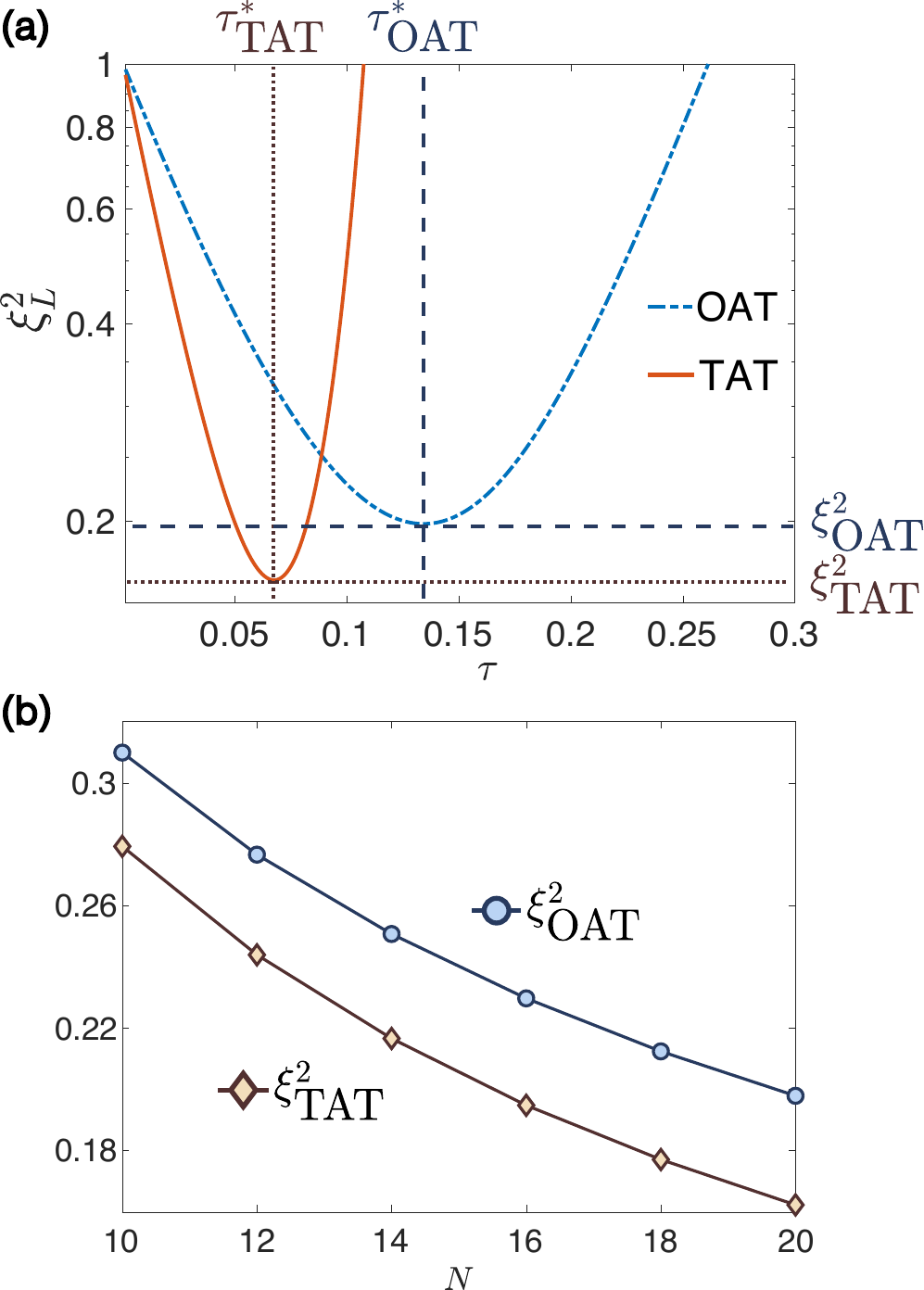}\\
  \caption{(a) Linear spin squeezing parameter $\xi^{2}_{L}$ as a function of the evolution time $\tau$ for the OAT and TAT with system size $N=20$. The dashed and dotted horizontal lines represent the minimum $\xi^{2}_{L}$ generated by the OAT and TAT, i.e., $\xi_{\text{OAT}}^{2} = 0.1979$ and $\xi_{\text{TAT}}^{2} = 0.1623$, respectively.  (b) The minimum $\xi^{2}_{L}$ generated by the OAT and TAT,  i.e., the $\xi^{2}_{\text{OAT}}$ and $\xi^{2}_{\text{TAT}}$ as a function of the system size $N$. }\label{s_2_1}
\end{figure}

\begin{figure}[t]
  \centering
  \includegraphics[width=1\linewidth]{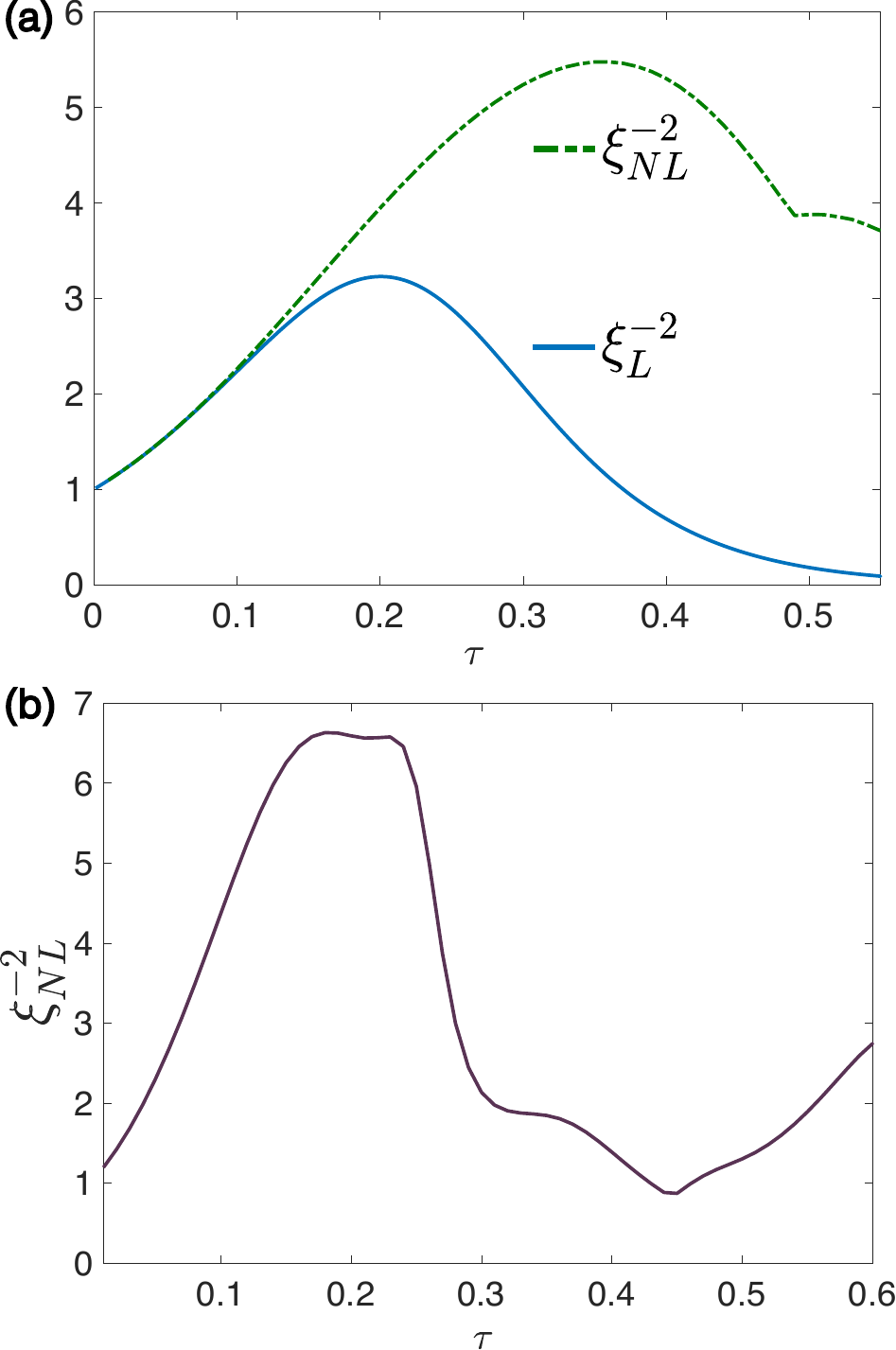}\\
  \caption{(a) For the OAT, the time evolution of both the linear spin squeezing parameter $\xi^{-2}_{L}$ and the second-order non-linear squeezing parameter $\xi^{-2}_{NL}$ for system size $N=10$. (b) For the TAT, the time evolution of $\xi^{-2}_{NL}$ with system size $N=10$. }\label{s_2_2}
\end{figure}

\section{Details of the optimization procedure of variational spin-squeezing algorithms}

In this section, we present details of the optimization procedure. We employ the BFGS method, as a gradient-based approach, to optimize the squeezing parameters. For the optimization of the function $f(\vec{x})$, with $\vec{x} = (x_{1},x_{2},...,x_{m})$ being the variational parameters, the gradients can be computed using forward finite differences, i.e., $\partial_{j}f(\vec{x}) \simeq [f(\vec{x} + \epsilon\vec{e}_{j}) - f(\vec{x})]/\epsilon$, where $\vec{e}_{j}$ is the unit vector with 1 as its $j$-th element and 0 otherwise, and $\epsilon$ denotes the finite-difference step size. Here, we chose $\epsilon = 10^{-8}$, which has been found to effectively suppress statistical fluctuations in our applications. For each step of the optimization procedure, in order to obtain the gradient, the number of repetitions for the calculation of $f(\vec{x})$ is $m$. 

It has been recognized that the BFGS method looks for local minima. To avoid the influence of local minima, for systems with $N=10$, we perform 1000 optimizations using randomly chosen initial parameters $\vec{x}_{0}$, and extract the minimum from 1000 optimization trajectories. 
The optimized variational parameter for $N=10$ is denoted as $\vec{x}_{\text{opt}}^{(N=10)}$. For larger system size $N=12$, we find that $\vec{x}_{\text{opt}}^{(N=10)}$ is an appropriate initial parameter for the optimization procedure. Here, we consider the PQC shown in Fig.~\ref{fig1}(b) with depth $p=5$ as an example. As shown in Fig.~\ref{fig_re_2}, with the initial parameter $\vec{x}_{0}^{(N=12)} = \vec{x}_{\text{opt}}^{(N=10)}$, the squeezing parameter $\xi_{L}^{2}$ tends to a lower value during the optimization, in comparison with 10 randomly chosen  $\vec{x}_{0}^{(N=12)}$. Thus, we adopt the initial parameter $\vec{x}_{0}^{(N)} = \vec{x}_{\text{opt}}^{(N-2)}$ to efficiently perform the optimization with system sizes $N\geq 12$.

\section{One-axis twisting and two-axis twisting}

In this section, we first briefly introduce the one-axis twisting (OAT) and TAT. The OAT and TAT Hamiltonian can be written as $\hat{H}_{\text{OAT}} = \hat{J}_{z}^{2}$ and $\hat{H}_{\text{TAT}} = \hat{J}_{x}^{2} - \hat{J}_{y}^{2}$, respectively~\cite{MA201189,PhysRevA.47.5138}. The OAT is the dynamical process described by the unitary evolution under $\hat{H}_{\text{OAT}}$, i.e., $\hat{U}_{\text{OAT}}(\tau) = \exp(-i\hat{H}_{\text{OAT}}\tau)$. For the OAT, the initial state is chosen as $|\psi_{0}\rangle_{\text{OAT}} = \otimes_{j=1}^{N}|+\rangle_{j}$, with $|+\rangle$ being the eigenstate of $\hat{\sigma}^{x}$ with the eigenvalue $+1$. The non-equilibrium state generated from the OAT is $|\psi(\tau)\rangle_{\text{OAT}} = \hat{U}_{\text{OAT}}(\tau)|\psi_{0}\rangle_{\text{OAT}}$. Similarly, for the TAT, the unitary evolution is $\hat{U}_{\text{TAT}}(\tau) = \exp(-i\hat{H}_{\text{TAT}}\tau)$, and $|\psi(\tau)\rangle_{\text{TAT}} = \hat{U}_{\text{TAT}}(\tau)|\psi_{0}\rangle_{\text{TAT}}$, with the chosen initial state for the TAT being $|\psi_{0}\rangle_{\text{TAT}} = \otimes_{j=1}^{N}|0\rangle_{j}$. 

We then can calculate the linear spin squeezing parameter defined in Eq.~(1) with $\hat{\textbf{S}}_{L}=(\hat{J}_{x},\hat{J}_{y},\hat{J}_{z})$, i.e., $\xi^{2}_{L}$, for the OAT states $|\psi(\tau)\rangle_{\text{OAT}}$ and the TAT states $|\psi(\tau)\rangle_{\text{TAT}}$. In Fig.~\ref{s_2_1}(a), we plot the $\xi^{2}_{L}$ as a function of the evolved time $\tau$ for both the OAT and TAT with system size $N=20$. We note that the minimum $\xi^{2}_{L}$, signaling the best spin-squeezed state generated by the OAT and TAT, can be achieved at the time $\tau_{\text{OAT}}^{*}$ and $\tau_{\text{TAT}}^{*}$, respectively. We denote the minimum $\xi^{2}_{L}$ generated by the OAT and TAT as $\xi^{2}_{\text{OAT}}$ and $\xi^{2}_{\text{TAT}}$, respectively. In Fig.~\ref{s_2_1}(b), we plot the $\xi^{2}_{\text{OAT}}$ and $\xi^{2}_{\text{TAT}}$ with different system sizes $N$, showing that the TAT can generate stronger squeezing than the OAT. 

Finally, for the OAT and TAT, we calculate the second-order non-linear squeezing parameter defined in the Eq.~(1) with $\hat{\textbf{S}}_{NL}=(\hat{J}_{x},\hat{J}_{y},\hat{J}_{z},\hat{J}_{x}^{2},\hat{J}_{y}^{2},\hat{J}_{z}^{2},\hat{J}_{xy}^{2},\hat{J}_{yz}^{2},\hat{J}_{zx}^{2})$. In Fig.~\ref{s_2_2}(a), we show the time evolution of $\xi^{-2}_{NL}$ in comparison with the $\xi^{-2}_{NL}$ in the OAT, demonstrating that $\xi^{-2}_{NL}\geq \xi^{-2}_{L}$. In Fig.~\ref{s_2_2}(b), we display the dynamics of $\xi^{-2}_{NL}$ in the TAT, showing that the TAT can also generate a state with larger $\xi^{-2}_{NL}$ than the OAT.

\begin{figure}[t]
  \centering
  \includegraphics[width=1\linewidth]{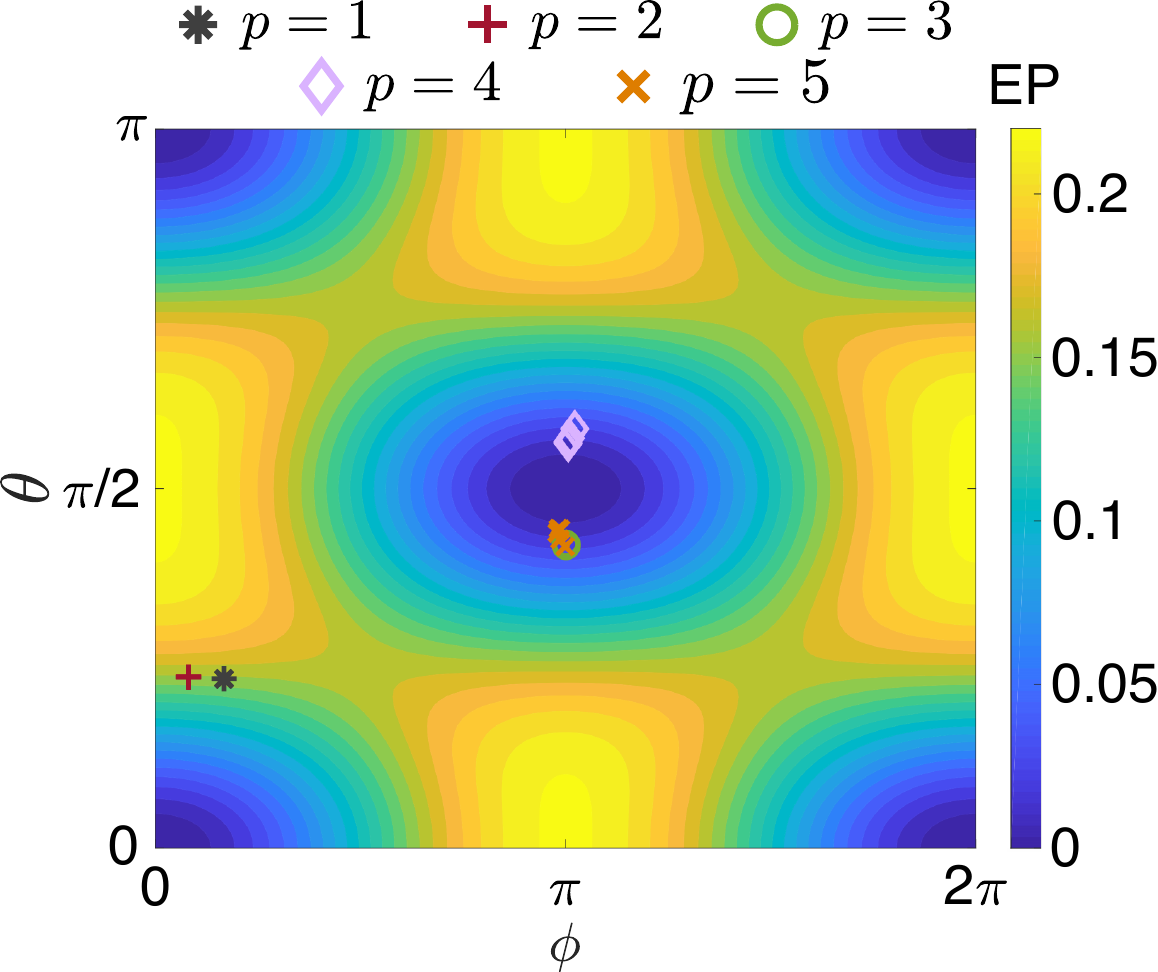}\\
  \caption{The entanglement power of FSIM gates, i.e., $\text{EP}[\text{FSIM}(\theta,\phi)]$, as a function of $\theta$ and $\phi$. The values of $\theta_{\text{opt}}$ and $\phi_{\text{opt}}$ in Table~\uppercase\expandafter{\romannumeral1} are marked. }\label{re_3}
\end{figure}

\begin{figure*}[]
  \centering
  \includegraphics[width=0.6\linewidth]{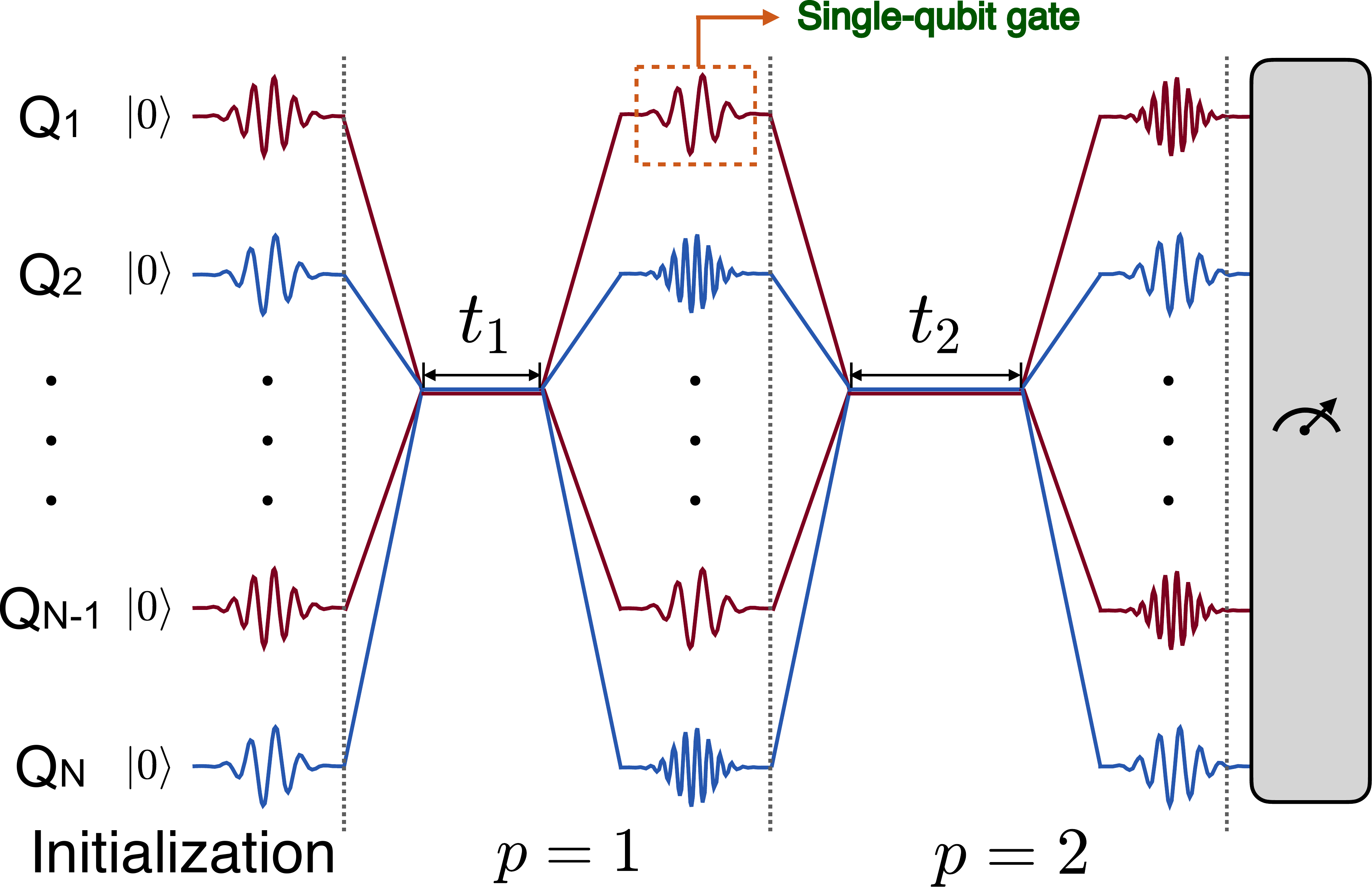}\\
  \caption{Schematic of the experimental pulse sequence corresponding to the PQC shown in Fig.~\ref{fig3}(a). After initialization, all qubits are tuned to the working point by a pulse lasting a time $t_{1}$, and the unitary evolution $\hat{U}(t_{1})$ can be realized. Next, all qubits are tuned to the idle point and single-qubit pulses are applied for single-qubit rotations. }\label{s_4_1}
\end{figure*}

\section{Details of the optimized FSIM gates}

In Table~\uppercase\expandafter{\romannumeral1}, we present the optimized angles $(\theta_{\text{opt}},\phi_{\text{opt}})$ of FSIM gates (\ref{fsim}) in the PQC shown in Fig.~\ref{fig1}(b) with different depth $p$ and system size $N$. For each $p$, one can see a weak dependence of $\theta_{\text{opt}}$ and  $\phi_{\text{opt}}$ on $N$. However, the values of $\theta_{\text{opt}}$ and  $\phi_{\text{opt}}$ are sensitive to the depth of the PQC $p$. For $p=1$ and $2$, $\theta_{\text{opt}}>\phi_{\text{opt}}$, while for $p\geq 3$,  $\theta_{\text{opt}}<\phi_{\text{opt}}$.

\begin{table}[htb]   
\begin{center}   
\begin{tabular}{c|c|c|c|c|c|c}   
\hline   \textbf{$p=1$} & $N=10$ & $N=12$ & $N=14$ & $N=16$ & $N=18$ & $N=20$ \\   
\Xhline{1px}    $\theta_{\text{opt}}$ & 0.7380 & 0.7380 & 0.7380 & 0.7380 & 0.7380 & 0.7380\\ 
\hline   $\phi_{\text{opt}}$ & 0.5256 & 0.5256 & 0.5256 & 0.5256 & 0.5256 & 0.5256\\    
\hline  
\hline \textbf{$p=2$} & $N=10$ & $N=12$ & $N=14$ & $N=16$ & $N=18$ & $N=20$ \\   
\Xhline{1px}   $\theta_{\text{opt}}$ & 0.7468 & 0.7474 & 0.7470 & 0.7470 & 0.7470 & 0.7470\\ 
\hline   $\phi_{\text{opt}}$ & 0.2542 & 0.2543 & 0.2544 & 0.2544 & 0.2544 & 0.2544\\    
\hline   
\hline \textbf{$p=3$} & $N=10$ & $N=12$ & $N=14$ & $N=16$ & $N=18$ & $N=20$ \\   
\Xhline{1px}   $\theta_{\text{opt}}$ & 1.3200 & 1.3242 & 1.3287 & 1.3287 & 1.3287 & 1.3288\\ 
\hline   $\phi_{\text{opt}}$ & 3.1449 & 3.1494 & 3.1408 & 3.1408 & 3.1408 & 3.1439\\    
\hline
\hline \textbf{$p=4$} & $N=10$ & $N=12$ & $N=14$ & $N=16$ & $N=18$ & $N=20$ \\   
\Xhline{1px}   $\theta_{\text{opt}}$ & 1.8320 & 1.7906 & 1.7746 & 1.7693 & 1.7683 & 1.7683\\ 
\hline   $\phi_{\text{opt}}$ & 3.2124 & 3.1775 & 3.1589 & 3.1614 & 3.1614 & 3.1614\\    
\hline
\hline \textbf{$p=5$} & $N=10$ & $N=12$ & $N=14$ & $N=16$ & $N=18$ & $N=20$ \\   
\Xhline{1px}   $\theta_{\text{opt}}$ & 1.3175 & 1.3482 & 1.3778 & 1.3928 & 1.3955 & 1.3955\\ 
\hline   $\phi_{\text{opt}}$ & 3.1379 & 3.0976 & 3.0768 & 3.1018 & 3.0865 & 3.0865\\    
\hline
\end{tabular}   
\caption{Optimized values of $\theta$ and $\phi$ of the FSIM gate for different values of $p$ and $N$.}  
\end{center}   
\label{t1} 
\end{table}

To quantify the entanglement of the FSIM gates defined by Eq.~(\ref{fsim}), we consider the entanglement power (EP)~\cite{PhysRevA.62.030301}, which is express as 
\begin{eqnarray}
\text{EP}(\hat{U}) = \overline{E(\hat{U}|\psi_{1}\rangle\otimes|\psi_{2}\rangle)},
\label{ep_def}
\end{eqnarray}
where $|\psi_{1}\rangle$ and $|\psi_{2}\rangle$ are single-qubit states uniformly sampled on the Bloch sphere, the top bar denotes the average over to all states $|\psi_{1}\rangle\otimes|\psi_{2}\rangle$, and $E(|\Psi\rangle) = 1 - \text{Tr}(\hat{\rho}_{1})$ ($\hat{\rho}_{1} = \text{Tr}_{2}|\Psi\rangle\langle\Psi|$) being the linear entropy.  The entanglement power of the FSIM gate $\text{FSIM}(\theta,\phi)$ can be analytically derived as $\text{EP}[\text{FSIM}(\theta,\phi)] = -\cos2\theta(\cos2\theta+\cos\phi)/18 + 1/6$. The maximum value of $\text{EP}[\text{FSIM}(\theta,\phi)]$ is $2/9$. We plot $\text{EP}[\text{FSIM}(\theta,\phi)]$ as a function of $\theta$ and $\phi$ in Fig.~\ref{re_3}. One can see that for all $p=1,2,...,5$, the optimized values $(\theta_{\text{opt}},\phi_{\text{opt}})$ do not lie in the region with maximum entanglement power. Consequently, the variational spin-squeezing algorithm with the PQC shown in Fig.~\ref{fig1}(b) does not require FSIM gates with large entanglement power. 

\section{Experimental realization of the global entanglers by using analog superconducting circuits}

In this section, we will present more details for the experimental realization of the global entanglers under the time evolution of the Hamiltonian $\hat{H}_{T} = \sum_{i=1}^{N} (\hat{\sigma}_{i}^{x}\hat{\sigma}_{i+1}^{x} + \hat{\sigma}_{i}^{y}\hat{\sigma}_{i+1}^{y})$ by using superconducting circuits. Actually, a loop of superconducting qubits coupled via a fixed capacitor can be described by the Bose-Hubbard model~\cite{sc5,sc6,sc9,sc10}
\begin{eqnarray}
\hat{H}_{\textrm{BH}} = &J& \sum_{i=1}^{N}(\hat{a}_{i}^{\dagger}\hat{a}_{i+1} + \textrm{H.c.}) \\ \nonumber
&+& \frac{U}{2}\sum_{i=1}^{N} \hat{n}_{i}(\hat{n}_{i} - 1) + \sum_{i=1}^{N}\mu_{i}\hat{n}_{i}, 
\label{h_bh}
\end{eqnarray}
where $\hat{n}_{i} = \hat{a}_{i}^{\dagger} \hat{a}_{i}$, $\hat{a}_{i}$ ($\hat{a}_{i}^{\dagger}$) denotes the bosonic annihilation (creation) operator, $J$ is the strength of the hopping interaction, $U$ is the nonlinear interaction, and $\mu_{i}$ is the chemical potential. 

In a typical superconducting quantum processor, the ratio $|U|/J$ is quite large. As an example, $|U|/2\pi \simeq 230$ MHz, $J/2\pi \simeq 12$ MHz, and $|U|/J \simeq 19$ for the quantum device in Ref.~\cite{sc6} .  In this situation, the Bose-Hubbard model Eq.~(\ref{h_bh}) is approximate to the hard-core limit, where the bosonic operators $\hat{a}_{i}$ and $\hat{a}_{i}^{\dagger}$ are replaced by the spin raising and lowering operators, respectively. The Hamiltonian with the hard-core limit (also known as the $XY$ model) can be written as 
\begin{eqnarray}
\hat{H}_{XY} = J \sum_{i=1}^{N}(\hat{\sigma}_{i}^{+}\hat{\sigma}_{i+1}^{-} + \textrm{H.c.}) + \sum_{i=1}^{N}\mu_{i}\hat{\sigma}_{i}^{+} \hat{\sigma}_{i}^{-}. 
\label{h_xx}
\end{eqnarray}
When we tune all qubits to the same working point (see Fig.~\ref{s_4_1}), the chemical potentials satisfy $\mu_{i}=0$, and the unitary evolution $\hat{U}(t) = \exp(-i\hat{H}_{XY}\tau) = \exp[-i\sum_{j=1}^{N}(\hat{\sigma}_{j}^{x}\hat{\sigma}_{j+1}^{x} + \hat{\sigma}_{j}^{y}\hat{\sigma}_{j+1}^{y} )t]$, with $t = J\tau/2$ can be experimentally realized. An experimental pulse sequence with control waveforms corresponding to the PQC shown in Fig.~\ref{fig3}(a) is depicted in  Fig.~\ref{s_4_1}.

\section{FSIM gates with coherent errors} 

\noindent The FSIM gate with coherent errors can be written as~\cite{coherent_error_OTOC}
\begin{align} 
\text{FSIM}_{\text{exp.}} &=& \text{FSIM}_{\text{exp.}}(\theta_{*},\phi_{*},\Delta_{+}, \Delta_{-}, \Delta_{-,\text{off}}) \\ 
& = &  
\begin{pmatrix}
1 & 0 & 0 & 0 \\
0 & u_{22} & u_{23} & 0 \\
0 & u_{32} & u_{33} & 0 \\
0 & 0 & 0 & u_{44} 
\end{pmatrix}, 
\label{fsim_error}
\end{align}
where 
\begin{eqnarray}
u_{22} = \exp\{i(\Delta_{+} + \Delta_{-})\}\cos\theta_{*}, 
\label{u22}
\end{eqnarray}
\begin{eqnarray}
u_{23} = -i\exp\{i(\Delta_{+} - \Delta_{-,\text{off}})\}\sin\theta_{*}, 
\label{u23}
\end{eqnarray}
\begin{eqnarray}
u_{32} = -i\exp\{i(\Delta_{+} + \Delta_{-,\text{off}})\}\sin\theta_{*}, 
\label{u32}
\end{eqnarray}
\begin{eqnarray}
u_{33} = \exp\{i(\Delta_{+} - \Delta_{-})\}\cos\theta_{*}, 
\label{u33}
\end{eqnarray}
\begin{eqnarray}
u_{44} = \exp\{i(2\Delta_{+}-\phi_{*})\}. 
\label{u44}
\end{eqnarray}

The influence of coherent errors can be analyzed by simulating the PQC with the FSIM gates described by Eq.~(\ref{fsim_error}). Taking the PQC based on the ALA shown in Fig.~\ref{fig1}(b) with $N=12$ and the degree of coherent error $r=0.005$ as an example, we can numerically generate a series of FSIM gates, i.e., $\text{FSIM}_{\text{exp.}}^{(n,m)} = \text{FSIM}_{\text{exp.}}(\theta_{*}^{(n,m)} ,\phi_{*}^{(n,m)} ,\Delta_{+}^{(n,m)} , \Delta_{-}^{(n,m)} , \Delta_{-,\text{off}}^{(n,m)} )$ as the FSIM gate between the $n$-th and $m$-th qubit (here $m=n+1$, for $n=1,2,...,11$, and $m=1$ for $n=12$), which satisfies $r=0.005$. It is noted that the parameters $(\theta_{*}^{(n,m)} ,\phi_{*}^{(n,m)} ,\Delta_{+}^{(n,m)} , \Delta_{-}^{(n,m)} , \Delta_{-,\text{off}}^{(n,m)} )$ are not necessarily identical for different $n$. After generating the FSIM gates with coherent errors, we can perform the optimization algorithm for the parameters in the single-rotations, and obtain the optimized squeezing parameter for the PQC with coherent errors.





\bibliography{reference_sss}

\begin{thebibliography}{76}%
\makeatletter
\providecommand \@ifxundefined [1]{%
 \@ifx{#1\undefined}
}%
\providecommand \@ifnum [1]{%
 \ifnum #1\expandafter \@firstoftwo
 \else \expandafter \@secondoftwo
 \fi
}%
\providecommand \@ifx [1]{%
 \ifx #1\expandafter \@firstoftwo
 \else \expandafter \@secondoftwo
 \fi
}%
\providecommand \natexlab [1]{#1}%
\providecommand \enquote  [1]{``#1''}%
\providecommand \bibnamefont  [1]{#1}%
\providecommand \bibfnamefont [1]{#1}%
\providecommand \citenamefont [1]{#1}%
\providecommand \href@noop [0]{\@secondoftwo}%
\providecommand \href [0]{\begingroup \@sanitize@url \@href}%
\providecommand \@href[1]{\@@startlink{#1}\@@href}%
\providecommand \@@href[1]{\endgroup#1\@@endlink}%
\providecommand \@sanitize@url [0]{\catcode `\\12\catcode `\$12\catcode
  `\&12\catcode `\#12\catcode `\^12\catcode `\_12\catcode `\%12\relax}%
\providecommand \@@startlink[1]{}%
\providecommand \@@endlink[0]{}%
\providecommand \url  [0]{\begingroup\@sanitize@url \@url }%
\providecommand \@url [1]{\endgroup\@href {#1}{\urlprefix }}%
\providecommand \urlprefix  [0]{URL }%
\providecommand \Eprint [0]{\href }%
\providecommand \doibase [0]{http://dx.doi.org/}%
\providecommand \selectlanguage [0]{\@gobble}%
\providecommand \bibinfo  [0]{\@secondoftwo}%
\providecommand \bibfield  [0]{\@secondoftwo}%
\providecommand \translation [1]{[#1]}%
\providecommand \BibitemOpen [0]{}%
\providecommand \bibitemStop [0]{}%
\providecommand \bibitemNoStop [0]{.\EOS\space}%
\providecommand \EOS [0]{\spacefactor3000\relax}%
\providecommand \BibitemShut  [1]{\csname bibitem#1\endcsname}%
\let\auto@bib@innerbib\@empty
\bibitem [{\citenamefont {Pezz\`e}\ \emph {et~al.}(2018)\citenamefont
  {Pezz\`e}, \citenamefont {Smerzi}, \citenamefont {Oberthaler}, \citenamefont
  {Schmied},\ and\ \citenamefont {Treutlein}}]{RevModPhys.90.035005}%
  \BibitemOpen
  \bibfield  {author} {\bibinfo {author} {\bibfnamefont {L.}~\bibnamefont
  {Pezz\`e}}, \bibinfo {author} {\bibfnamefont {A.}~\bibnamefont {Smerzi}},
  \bibinfo {author} {\bibfnamefont {M.~K.}\ \bibnamefont {Oberthaler}},
  \bibinfo {author} {\bibfnamefont {R.}~\bibnamefont {Schmied}}, \ and\
  \bibinfo {author} {\bibfnamefont {P.}~\bibnamefont {Treutlein}},\ }\bibfield
  {title} {\enquote {\bibinfo {title} {Quantum metrology with nonclassical
  states of atomic ensembles},}\ }\href {\doibase 10.1103/RevModPhys.90.035005}
  {\bibfield  {journal} {\bibinfo  {journal} {Rev. Mod. Phys.}\ }\textbf
  {\bibinfo {volume} {90}},\ \bibinfo {pages} {035005} (\bibinfo {year}
  {2018})}\BibitemShut {NoStop}%
\bibitem [{\citenamefont {\emph{et~al.}}(2011)}]{MA201189}%
  \BibitemOpen
  \bibfield  {author} {\bibinfo {author} {\bibfnamefont {J.~Ma}\ \bibnamefont
  {\emph{et~al.}}},\ }\bibfield  {title} {\enquote {\bibinfo {title} {Quantum
  spin squeezing},}\ }\href {\doibase
  https://doi.org/10.1016/j.physrep.2011.08.003} {\bibfield  {journal}
  {\bibinfo  {journal} {Phys. Rep.}\ }\textbf {\bibinfo {volume} {509}},\
  \bibinfo {pages} {89--165} (\bibinfo {year} {2011})}\BibitemShut {NoStop}%
\bibitem [{\citenamefont {Giovannetti}\ \emph {et~al.}(2011)\citenamefont
  {Giovannetti}, \citenamefont {Lloyd},\ and\ \citenamefont {Maccone}}]{qm1}%
  \BibitemOpen
  \bibfield  {author} {\bibinfo {author} {\bibfnamefont {V.}~\bibnamefont
  {Giovannetti}}, \bibinfo {author} {\bibfnamefont {S.}~\bibnamefont {Lloyd}},
  \ and\ \bibinfo {author} {\bibfnamefont {L.}~\bibnamefont {Maccone}},\
  }\bibfield  {title} {\enquote {\bibinfo {title} {Advances in quantum
  metrology},}\ }\href {\doibase doi.org/10.1038/nphoton.2011.35} {\bibfield
  {journal} {\bibinfo  {journal} {Nat. Photonics}\ }\textbf {\bibinfo {volume}
  {5}},\ \bibinfo {pages} {222--229} (\bibinfo {year} {2011})}\BibitemShut
  {NoStop}%
\bibitem [{\citenamefont {Gross}\ \emph {et~al.}(2010)\citenamefont {Gross},
  \citenamefont {Zibold}, \citenamefont {Nicklas}, \citenamefont {Est\`eve},\
  and\ \citenamefont {Oberthaler}}]{qm2}%
  \BibitemOpen
  \bibfield  {author} {\bibinfo {author} {\bibfnamefont {C.}~\bibnamefont
  {Gross}}, \bibinfo {author} {\bibfnamefont {T.}~\bibnamefont {Zibold}},
  \bibinfo {author} {\bibfnamefont {E.}~\bibnamefont {Nicklas}}, \bibinfo
  {author} {\bibfnamefont {J.}~\bibnamefont {Est\`eve}}, \ and\ \bibinfo
  {author} {\bibfnamefont {M.~K.}\ \bibnamefont {Oberthaler}},\ }\bibfield
  {title} {\enquote {\bibinfo {title} {Nonlinear atom interferometer surpasses
  classical precision limit},}\ }\href {\doibase doi.org/10.1038/nature08919}
  {\bibfield  {journal} {\bibinfo  {journal} {Nature}\ }\textbf {\bibinfo
  {volume} {464}},\ \bibinfo {pages} {1165--1169} (\bibinfo {year}
  {2010})}\BibitemShut {NoStop}%
\bibitem [{\citenamefont {Riedel}\ \emph {et~al.}(2010)\citenamefont {Riedel},
  \citenamefont {B\"ohi}, \citenamefont {Li}, \citenamefont {H\"ansch},
  \citenamefont {Sinatra},\ and\ \citenamefont {Treutlein}}]{qm3}%
  \BibitemOpen
  \bibfield  {author} {\bibinfo {author} {\bibfnamefont {M.~F.}\ \bibnamefont
  {Riedel}}, \bibinfo {author} {\bibfnamefont {P.}~\bibnamefont {B\"ohi}},
  \bibinfo {author} {\bibfnamefont {Y.}~\bibnamefont {Li}}, \bibinfo {author}
  {\bibfnamefont {T.~W.}\ \bibnamefont {H\"ansch}}, \bibinfo {author}
  {\bibfnamefont {A.}~\bibnamefont {Sinatra}}, \ and\ \bibinfo {author}
  {\bibfnamefont {P.}~\bibnamefont {Treutlein}},\ }\bibfield  {title} {\enquote
  {\bibinfo {title} {Atom-chip-based generation of entanglement for quantum
  metrology},}\ }\href {\doibase doi.org/10.1038/nature08988} {\bibfield
  {journal} {\bibinfo  {journal} {Nature}\ }\textbf {\bibinfo {volume} {464}},\
  \bibinfo {pages} {1170--1173} (\bibinfo {year} {2010})}\BibitemShut {NoStop}%
\bibitem [{\citenamefont {Wolfgramm}\ \emph {et~al.}(2010)\citenamefont
  {Wolfgramm}, \citenamefont {Cer\`e}, \citenamefont {Beduini}, \citenamefont
  {Predojevi\ifmmode~\acute{c}\else \'{c}\fi{}}, \citenamefont {Koschorreck},\
  and\ \citenamefont {Mitchell}}]{PhysRevLett.105.053601}%
  \BibitemOpen
  \bibfield  {author} {\bibinfo {author} {\bibfnamefont {F.}~\bibnamefont
  {Wolfgramm}}, \bibinfo {author} {\bibfnamefont {A.}~\bibnamefont {Cer\`e}},
  \bibinfo {author} {\bibfnamefont {F.~A.}\ \bibnamefont {Beduini}}, \bibinfo
  {author} {\bibfnamefont {A.}~\bibnamefont {Predojevi\ifmmode~\acute{c}\else
  \'{c}\fi{}}}, \bibinfo {author} {\bibfnamefont {M.}~\bibnamefont
  {Koschorreck}}, \ and\ \bibinfo {author} {\bibfnamefont {M.~W.}\ \bibnamefont
  {Mitchell}},\ }\bibfield  {title} {\enquote {\bibinfo {title} {Squeezed-light
  optical magnetometry},}\ }\href {\doibase 10.1103/PhysRevLett.105.053601}
  {\bibfield  {journal} {\bibinfo  {journal} {Phys. Rev. Lett.}\ }\textbf
  {\bibinfo {volume} {105}},\ \bibinfo {pages} {053601} (\bibinfo {year}
  {2010})}\BibitemShut {NoStop}%
\bibitem [{\citenamefont {Backes~\emph{et~al.}}(2021)}]{Backes:2021wd}%
  \BibitemOpen
  \bibfield  {author} {\bibinfo {author} {\bibfnamefont {K.~M.}\ \bibnamefont
  {Backes~\emph{et~al.}}},\ }\bibfield  {title} {\enquote {\bibinfo {title} {A
  quantum enhanced search for dark matter axions},}\ }\href {\doibase
  10.1038/s41586-021-03226-7} {\bibfield  {journal} {\bibinfo  {journal}
  {Nature}\ }\textbf {\bibinfo {volume} {590}},\ \bibinfo {pages} {238--242}
  (\bibinfo {year} {2021})}\BibitemShut {NoStop}%
\bibitem [{\citenamefont {Huelga}\ \emph {et~al.}(1997)\citenamefont {Huelga},
  \citenamefont {Macchiavello}, \citenamefont {Pellizzari}, \citenamefont
  {Ekert}, \citenamefont {Plenio},\ and\ \citenamefont
  {Cirac}}]{PhysRevLett.79.3865}%
  \BibitemOpen
  \bibfield  {author} {\bibinfo {author} {\bibfnamefont {S.~F.}\ \bibnamefont
  {Huelga}}, \bibinfo {author} {\bibfnamefont {C.}~\bibnamefont
  {Macchiavello}}, \bibinfo {author} {\bibfnamefont {T.}~\bibnamefont
  {Pellizzari}}, \bibinfo {author} {\bibfnamefont {A.~K.}\ \bibnamefont
  {Ekert}}, \bibinfo {author} {\bibfnamefont {M.~B.}\ \bibnamefont {Plenio}}, \
  and\ \bibinfo {author} {\bibfnamefont {J.~I.}\ \bibnamefont {Cirac}},\
  }\bibfield  {title} {\enquote {\bibinfo {title} {Improvement of frequency
  standards with quantum entanglement},}\ }\href {\doibase
  10.1103/PhysRevLett.79.3865} {\bibfield  {journal} {\bibinfo  {journal}
  {Phys. Rev. Lett.}\ }\textbf {\bibinfo {volume} {79}},\ \bibinfo {pages}
  {3865--3868} (\bibinfo {year} {1997})}\BibitemShut {NoStop}%
\bibitem [{\citenamefont {Wang~\emph{et~al.}}(2010)}]{PhysRevA.81.022106}%
  \BibitemOpen
  \bibfield  {author} {\bibinfo {author} {\bibfnamefont {X.}~\bibnamefont
  {Wang~\emph{et~al.}}},\ }\bibfield  {title} {\enquote {\bibinfo {title}
  {Sudden vanishing of spin squeezing under decoherence},}\ }\href {\doibase
  10.1103/PhysRevA.81.022106} {\bibfield  {journal} {\bibinfo  {journal} {Phys.
  Rev. A}\ }\textbf {\bibinfo {volume} {81}},\ \bibinfo {pages} {022106}
  (\bibinfo {year} {2010})}\BibitemShut {NoStop}%
\bibitem [{\citenamefont {Kitagawa}\ and\ \citenamefont
  {Ueda}(1993)}]{PhysRevA.47.5138}%
  \BibitemOpen
  \bibfield  {author} {\bibinfo {author} {\bibfnamefont {M.}~\bibnamefont
  {Kitagawa}}\ and\ \bibinfo {author} {\bibfnamefont {M.}~\bibnamefont
  {Ueda}},\ }\bibfield  {title} {\enquote {\bibinfo {title} {Squeezed spin
  states},}\ }\href {\doibase 10.1103/PhysRevA.47.5138} {\bibfield  {journal}
  {\bibinfo  {journal} {Phys. Rev. A}\ }\textbf {\bibinfo {volume} {47}},\
  \bibinfo {pages} {5138--5143} (\bibinfo {year} {1993})}\BibitemShut {NoStop}%
\bibitem [{\citenamefont {Sebby-Strabley}\ \emph {et~al.}(2007)\citenamefont
  {Sebby-Strabley}, \citenamefont {Brown}, \citenamefont {Anderlini},
  \citenamefont {Lee}, \citenamefont {Phillips}, \citenamefont {Porto},\ and\
  \citenamefont {Johnson}}]{PhysRevLett.98.200405}%
  \BibitemOpen
  \bibfield  {author} {\bibinfo {author} {\bibfnamefont {J.}~\bibnamefont
  {Sebby-Strabley}}, \bibinfo {author} {\bibfnamefont {B.~L.}\ \bibnamefont
  {Brown}}, \bibinfo {author} {\bibfnamefont {M.}~\bibnamefont {Anderlini}},
  \bibinfo {author} {\bibfnamefont {P.~J.}\ \bibnamefont {Lee}}, \bibinfo
  {author} {\bibfnamefont {W.~D.}\ \bibnamefont {Phillips}}, \bibinfo {author}
  {\bibfnamefont {J.~V.}\ \bibnamefont {Porto}}, \ and\ \bibinfo {author}
  {\bibfnamefont {P.~R.}\ \bibnamefont {Johnson}},\ }\bibfield  {title}
  {\enquote {\bibinfo {title} {Preparing and probing atomic number states with
  an atom interferometer},}\ }\href {\doibase 10.1103/PhysRevLett.98.200405}
  {\bibfield  {journal} {\bibinfo  {journal} {Phys. Rev. Lett.}\ }\textbf
  {\bibinfo {volume} {98}},\ \bibinfo {pages} {200405} (\bibinfo {year}
  {2007})}\BibitemShut {NoStop}%
\bibitem [{\citenamefont {Muessel}\ \emph {et~al.}(2015)\citenamefont
  {Muessel}, \citenamefont {Strobel}, \citenamefont {Linnemann}, \citenamefont
  {Zibold}, \citenamefont {Juli\'a-D\'{\i}az},\ and\ \citenamefont
  {Oberthaler}}]{PhysRevA.92.023603}%
  \BibitemOpen
  \bibfield  {author} {\bibinfo {author} {\bibfnamefont {W.}~\bibnamefont
  {Muessel}}, \bibinfo {author} {\bibfnamefont {H.}~\bibnamefont {Strobel}},
  \bibinfo {author} {\bibfnamefont {D.}~\bibnamefont {Linnemann}}, \bibinfo
  {author} {\bibfnamefont {T.}~\bibnamefont {Zibold}}, \bibinfo {author}
  {\bibfnamefont {B.}~\bibnamefont {Juli\'a-D\'{\i}az}}, \ and\ \bibinfo
  {author} {\bibfnamefont {M.~K.}\ \bibnamefont {Oberthaler}},\ }\bibfield
  {title} {\enquote {\bibinfo {title} {Twist-and-turn spin squeezing in
  {B}ose-{E}instein condensates},}\ }\href {\doibase
  10.1103/PhysRevA.92.023603} {\bibfield  {journal} {\bibinfo  {journal} {Phys.
  Rev. A}\ }\textbf {\bibinfo {volume} {92}},\ \bibinfo {pages} {023603}
  (\bibinfo {year} {2015})}\BibitemShut {NoStop}%
\bibitem [{\citenamefont {S{\o}rensen}\ \emph {et~al.}(2001)\citenamefont
  {S{\o}rensen}, \citenamefont {Duan}, \citenamefont {Cirac},\ and\
  \citenamefont {Zoller}}]{qm4}%
  \BibitemOpen
  \bibfield  {author} {\bibinfo {author} {\bibfnamefont {A.}~\bibnamefont
  {S{\o}rensen}}, \bibinfo {author} {\bibfnamefont {L.-M.}\ \bibnamefont
  {Duan}}, \bibinfo {author} {\bibfnamefont {J.~I.}\ \bibnamefont {Cirac}}, \
  and\ \bibinfo {author} {\bibfnamefont {P.}~\bibnamefont {Zoller}},\
  }\bibfield  {title} {\enquote {\bibinfo {title} {Many-particle entanglement
  with {B}ose--{E}instein condensates},}\ }\href {\doibase
  doi.org/10.1038/35051038} {\bibfield  {journal} {\bibinfo  {journal}
  {Nature}\ }\textbf {\bibinfo {volume} {409}},\ \bibinfo {pages} {63--66}
  (\bibinfo {year} {2001})}\BibitemShut {NoStop}%
\bibitem [{\citenamefont {Strobel~\emph {et~al.}}(2014)}]{qm5}%
  \BibitemOpen
  \bibfield  {author} {\bibinfo {author} {\bibfnamefont {H.}~\bibnamefont
  {Strobel~\emph {et~al.}}},\ }\bibfield  {title} {\enquote {\bibinfo {title}
  {Fisher information and entanglement of non-gaussian spin states},}\ }\href
  {\doibase 10.1126/science.1250147} {\bibfield  {journal} {\bibinfo  {journal}
  {Science}\ }\textbf {\bibinfo {volume} {345}},\ \bibinfo {pages} {424--427}
  (\bibinfo {year} {2014})}\BibitemShut {NoStop}%
\bibitem [{\citenamefont {Bohnet~\emph {et~al.}}(2016)}]{ti_oat}%
  \BibitemOpen
  \bibfield  {author} {\bibinfo {author} {\bibfnamefont {J.~G.}\ \bibnamefont
  {Bohnet~\emph {et~al.}}},\ }\bibfield  {title} {\enquote {\bibinfo {title}
  {Quantum spin dynamics and entanglement generation with hundreds of trapped
  ions},}\ }\href {\doibase 10.1126/science.aad9958} {\bibfield  {journal}
  {\bibinfo  {journal} {Science}\ }\textbf {\bibinfo {volume} {352}},\ \bibinfo
  {pages} {1297--1301} (\bibinfo {year} {2016})}\BibitemShut {NoStop}%
\bibitem [{\citenamefont {Eckner}\ \emph {et~al.}(2023)\citenamefont {Eckner},
  \citenamefont {Darkwah~Oppong}, \citenamefont {Cao}, \citenamefont {Young},
  \citenamefont {Milner}, \citenamefont {Robinson}, \citenamefont {Ye},\ and\
  \citenamefont {Kaufman}}]{2023arXiv230308078E}%
  \BibitemOpen
  \bibfield  {author} {\bibinfo {author} {\bibfnamefont {W.~J.}\ \bibnamefont
  {Eckner}}, \bibinfo {author} {\bibfnamefont {N.}~\bibnamefont
  {Darkwah~Oppong}}, \bibinfo {author} {\bibfnamefont {A.}~\bibnamefont {Cao}},
  \bibinfo {author} {\bibfnamefont {A.~W.}\ \bibnamefont {Young}}, \bibinfo
  {author} {\bibfnamefont {W.~R.}\ \bibnamefont {Milner}}, \bibinfo {author}
  {\bibfnamefont {John~M.}\ \bibnamefont {Robinson}}, \bibinfo {author}
  {\bibfnamefont {J.}~\bibnamefont {Ye}}, \ and\ \bibinfo {author}
  {\bibfnamefont {A.~M.}\ \bibnamefont {Kaufman}},\ }\bibfield  {title}
  {\enquote {\bibinfo {title} {Realizing spin squeezing with rydberg
  interactions in an optical clock},}\ }\href {\doibase
  10.1038/s41586-023-06360-6} {\bibfield  {journal} {\bibinfo  {journal}
  {Nature}\ }\textbf {\bibinfo {volume} {621}},\ \bibinfo {pages} {734--739}
  (\bibinfo {year} {2023})}\BibitemShut {NoStop}%
\bibitem [{\citenamefont {Song~\emph {et~al.}}(2019)}]{qc_all_1}%
  \BibitemOpen
  \bibfield  {author} {\bibinfo {author} {\bibfnamefont {C.}~\bibnamefont
  {Song~\emph {et~al.}}},\ }\bibfield  {title} {\enquote {\bibinfo {title}
  {Generation of multicomponent atomic {S}chr\"{o}dinger cat states of up to 20
  qubits},}\ }\href {\doibase 10.1126/science.aay0600} {\bibfield  {journal}
  {\bibinfo  {journal} {Science}\ }\textbf {\bibinfo {volume} {365}},\ \bibinfo
  {pages} {574--577} (\bibinfo {year} {2019})}\BibitemShut {NoStop}%
\bibitem [{\citenamefont {Xu~\emph {et~al.}}(2020)}]{qc_all_2}%
  \BibitemOpen
  \bibfield  {author} {\bibinfo {author} {\bibfnamefont {K.}~\bibnamefont
  {Xu~\emph {et~al.}}},\ }\bibfield  {title} {\enquote {\bibinfo {title}
  {Probing dynamical phase transitions with a superconducting quantum
  simulator},}\ }\href {\doibase 10.1126/sciadv.aba4935} {\bibfield  {journal}
  {\bibinfo  {journal} {Sci. Adv.}\ }\textbf {\bibinfo {volume} {6}},\ \bibinfo
  {pages} {eaba4935} (\bibinfo {year} {2020})}\BibitemShut {NoStop}%
\bibitem [{\citenamefont {Xu~\emph {et~al.}}(2022)}]{PhysRevLett.128.150501}%
  \BibitemOpen
  \bibfield  {author} {\bibinfo {author} {\bibfnamefont {K.}~\bibnamefont
  {Xu~\emph {et~al.}}},\ }\bibfield  {title} {\enquote {\bibinfo {title}
  {Metrological characterization of non-{G}aussian entangled states of
  superconducting qubits},}\ }\href {\doibase 10.1103/PhysRevLett.128.150501}
  {\bibfield  {journal} {\bibinfo  {journal} {Phys. Rev. Lett.}\ }\textbf
  {\bibinfo {volume} {128}},\ \bibinfo {pages} {150501} (\bibinfo {year}
  {2022})}\BibitemShut {NoStop}%
\bibitem [{\citenamefont {Liu}\ \emph {et~al.}(2011)\citenamefont {Liu},
  \citenamefont {Xu}, \citenamefont {Jin},\ and\ \citenamefont {You}}]{tat_1}%
  \BibitemOpen
  \bibfield  {author} {\bibinfo {author} {\bibfnamefont {Y.~C.}\ \bibnamefont
  {Liu}}, \bibinfo {author} {\bibfnamefont {Z.~F.}\ \bibnamefont {Xu}},
  \bibinfo {author} {\bibfnamefont {G.~R.}\ \bibnamefont {Jin}}, \ and\
  \bibinfo {author} {\bibfnamefont {L.}~\bibnamefont {You}},\ }\bibfield
  {title} {\enquote {\bibinfo {title} {Spin squeezing: Transforming one-axis
  twisting into two-axis twisting},}\ }\href {\doibase
  10.1103/PhysRevLett.107.013601} {\bibfield  {journal} {\bibinfo  {journal}
  {Phys. Rev. Lett.}\ }\textbf {\bibinfo {volume} {107}},\ \bibinfo {pages}
  {013601} (\bibinfo {year} {2011})}\BibitemShut {NoStop}%
\bibitem [{\citenamefont {Huang}\ \emph {et~al.}(2015)\citenamefont {Huang},
  \citenamefont {Zhang}, \citenamefont {Zou}, \citenamefont {Zou},\ and\
  \citenamefont {Guo}}]{tat_2}%
  \BibitemOpen
  \bibfield  {author} {\bibinfo {author} {\bibfnamefont {W.}~\bibnamefont
  {Huang}}, \bibinfo {author} {\bibfnamefont {Y.-L.}\ \bibnamefont {Zhang}},
  \bibinfo {author} {\bibfnamefont {C.-L.}\ \bibnamefont {Zou}}, \bibinfo
  {author} {\bibfnamefont {X.-B.}\ \bibnamefont {Zou}}, \ and\ \bibinfo
  {author} {\bibfnamefont {G.-C.}\ \bibnamefont {Guo}},\ }\bibfield  {title}
  {\enquote {\bibinfo {title} {Two-axis spin squeezing of two-component
  {B}ose-{E}instein condensates via continuous driving},}\ }\href {\doibase
  10.1103/PhysRevA.91.043642} {\bibfield  {journal} {\bibinfo  {journal} {Phys.
  Rev. A}\ }\textbf {\bibinfo {volume} {91}},\ \bibinfo {pages} {043642}
  (\bibinfo {year} {2015})}\BibitemShut {NoStop}%
\bibitem [{\citenamefont {Hern\'andez~Yanes}\ \emph {et~al.}(2022)\citenamefont
  {Hern\'andez~Yanes}, \citenamefont {P\l{}odzie\ifmmode~\acute{n}\else
  \'{n}\fi{}}, \citenamefont {Mackoit Sinkevi\ifmmode \check{c}\else
  \v{c}\fi{}ien\ifmmode~\dot{e}\else \.{e}\fi{}}, \citenamefont
  {\ifmmode~\check{Z}\else \v{Z}\fi{}labys}, \citenamefont
  {Juzeli\ifmmode~\bar{u}\else \={u}\fi{}nas},\ and\ \citenamefont
  {Witkowska}}]{tat_3}%
  \BibitemOpen
  \bibfield  {author} {\bibinfo {author} {\bibfnamefont {T.}~\bibnamefont
  {Hern\'andez~Yanes}}, \bibinfo {author} {\bibfnamefont {M.}~\bibnamefont
  {P\l{}odzie\ifmmode~\acute{n}\else \'{n}\fi{}}}, \bibinfo {author}
  {\bibfnamefont {M.}~\bibnamefont {Mackoit Sinkevi\ifmmode \check{c}\else
  \v{c}\fi{}ien\ifmmode~\dot{e}\else \.{e}\fi{}}}, \bibinfo {author}
  {\bibfnamefont {G.}~\bibnamefont {\ifmmode~\check{Z}\else \v{Z}\fi{}labys}},
  \bibinfo {author} {\bibfnamefont {G.}~\bibnamefont
  {Juzeli\ifmmode~\bar{u}\else \={u}\fi{}nas}}, \ and\ \bibinfo {author}
  {\bibfnamefont {E.}~\bibnamefont {Witkowska}},\ }\bibfield  {title} {\enquote
  {\bibinfo {title} {One- and two-axis squeezing via laser coupling in an
  atomic {F}ermi-{H}ubbard model},}\ }\href {\doibase
  10.1103/PhysRevLett.129.090403} {\bibfield  {journal} {\bibinfo  {journal}
  {Phys. Rev. Lett.}\ }\textbf {\bibinfo {volume} {129}},\ \bibinfo {pages}
  {090403} (\bibinfo {year} {2022})}\BibitemShut {NoStop}%
\bibitem [{\citenamefont {Wang}\ \emph {et~al.}(2017)\citenamefont {Wang},
  \citenamefont {Qu}, \citenamefont {Li}, \citenamefont {Bao}, \citenamefont
  {Vuleti\ifmmode~\acute{c}\else \'{c}\fi{}},\ and\ \citenamefont
  {Xiao}}]{tat_4}%
  \BibitemOpen
  \bibfield  {author} {\bibinfo {author} {\bibfnamefont {M.}~\bibnamefont
  {Wang}}, \bibinfo {author} {\bibfnamefont {W.}~\bibnamefont {Qu}}, \bibinfo
  {author} {\bibfnamefont {P.}~\bibnamefont {Li}}, \bibinfo {author}
  {\bibfnamefont {H.}~\bibnamefont {Bao}}, \bibinfo {author} {\bibfnamefont
  {V.}~\bibnamefont {Vuleti\ifmmode~\acute{c}\else \'{c}\fi{}}}, \ and\
  \bibinfo {author} {\bibfnamefont {Y.}~\bibnamefont {Xiao}},\ }\bibfield
  {title} {\enquote {\bibinfo {title} {Two-axis-twisting spin squeezing by
  multipass quantum erasure},}\ }\href {\doibase 10.1103/PhysRevA.96.013823}
  {\bibfield  {journal} {\bibinfo  {journal} {Phys. Rev. A}\ }\textbf {\bibinfo
  {volume} {96}},\ \bibinfo {pages} {013823} (\bibinfo {year}
  {2017})}\BibitemShut {NoStop}%
\bibitem [{\citenamefont {Borregaard}\ \emph {et~al.}(2017)\citenamefont
  {Borregaard}, \citenamefont {Davis}, \citenamefont {Bentsen}, \citenamefont
  {Schleier-Smith},\ and\ \citenamefont {S{\o}rensen}}]{tat_5}%
  \BibitemOpen
  \bibfield  {author} {\bibinfo {author} {\bibfnamefont {J}~\bibnamefont
  {Borregaard}}, \bibinfo {author} {\bibfnamefont {E~J}\ \bibnamefont {Davis}},
  \bibinfo {author} {\bibfnamefont {G~S}\ \bibnamefont {Bentsen}}, \bibinfo
  {author} {\bibfnamefont {M~H}\ \bibnamefont {Schleier-Smith}}, \ and\
  \bibinfo {author} {\bibfnamefont {A~S}\ \bibnamefont {S{\o}rensen}},\
  }\bibfield  {title} {\enquote {\bibinfo {title} {One- and two-axis squeezing
  of atomic ensembles in optical cavities},}\ }\href {\doibase
  10.1088/1367-2630/aa8438} {\bibfield  {journal} {\bibinfo  {journal} {New J.
  Phys.}\ }\textbf {\bibinfo {volume} {19}},\ \bibinfo {pages} {093021}
  (\bibinfo {year} {2017})}\BibitemShut {NoStop}%
\bibitem [{\citenamefont {Perlin}\ \emph {et~al.}(2020)\citenamefont {Perlin},
  \citenamefont {Qu},\ and\ \citenamefont {Rey}}]{PhysRevLett.125.223401}%
  \BibitemOpen
  \bibfield  {author} {\bibinfo {author} {\bibfnamefont {M.~A.}\ \bibnamefont
  {Perlin}}, \bibinfo {author} {\bibfnamefont {C.}~\bibnamefont {Qu}}, \ and\
  \bibinfo {author} {\bibfnamefont {A.~M.}\ \bibnamefont {Rey}},\ }\bibfield
  {title} {\enquote {\bibinfo {title} {Spin squeezing with short-range
  spin-exchange interactions},}\ }\href {\doibase
  10.1103/PhysRevLett.125.223401} {\bibfield  {journal} {\bibinfo  {journal}
  {Phys. Rev. Lett.}\ }\textbf {\bibinfo {volume} {125}},\ \bibinfo {pages}
  {223401} (\bibinfo {year} {2020})}\BibitemShut {NoStop}%
\bibitem [{\citenamefont {{Bornet \emph{et~al.}}}(2023)}]{2023arXiv230308053B}%
  \BibitemOpen
  \bibfield  {author} {\bibinfo {author} {\bibfnamefont {G.}~\bibnamefont
  {{Bornet \emph{et~al.}}}},\ }\bibfield  {title} {\enquote {\bibinfo {title}
  {Scalable spin squeezing in a dipolar rydberg atom array},}\ }\href {\doibase
  10.1038/s41586-023-06414-9} {\bibfield  {journal} {\bibinfo  {journal}
  {Nature}\ }\textbf {\bibinfo {volume} {621}},\ \bibinfo {pages} {728--733}
  (\bibinfo {year} {2023})}\BibitemShut {NoStop}%
\bibitem [{\citenamefont {{Block \emph{et~al.}}}(2023)}]{2023arXiv230109636B}%
  \BibitemOpen
  \bibfield  {author} {\bibinfo {author} {\bibfnamefont {Maxwell}\ \bibnamefont
  {{Block \emph{et~al.}}}},\ }\bibfield  {title} {\enquote {\bibinfo {title}
  {{A Universal Theory of Spin Squeezing}},}\ }\href {\doibase
  10.48550/arXiv.2301.09636} {\bibfield  {journal} {\bibinfo  {journal} {arXiv
  e-prints}\ ,\ \bibinfo {pages} {arXiv:2301.09636}} (\bibinfo {year}
  {2023})},\ \Eprint {http://arxiv.org/abs/2301.09636} {arXiv:2301.09636
  [quant-ph]} \BibitemShut {NoStop}%
\bibitem [{\citenamefont {Gil}\ \emph {et~al.}(2014)\citenamefont {Gil},
  \citenamefont {Mukherjee}, \citenamefont {Bridge}, \citenamefont {Jones},\
  and\ \citenamefont {Pohl}}]{PhysRevLett.112.103601}%
  \BibitemOpen
  \bibfield  {author} {\bibinfo {author} {\bibfnamefont {L.~I.~R.}\
  \bibnamefont {Gil}}, \bibinfo {author} {\bibfnamefont {R.}~\bibnamefont
  {Mukherjee}}, \bibinfo {author} {\bibfnamefont {E.~M.}\ \bibnamefont
  {Bridge}}, \bibinfo {author} {\bibfnamefont {M.~P.~A.}\ \bibnamefont
  {Jones}}, \ and\ \bibinfo {author} {\bibfnamefont {T.}~\bibnamefont {Pohl}},\
  }\bibfield  {title} {\enquote {\bibinfo {title} {Spin squeezing in a
  {R}ydberg lattice clock},}\ }\href {\doibase 10.1103/PhysRevLett.112.103601}
  {\bibfield  {journal} {\bibinfo  {journal} {Phys. Rev. Lett.}\ }\textbf
  {\bibinfo {volume} {112}},\ \bibinfo {pages} {103601} (\bibinfo {year}
  {2014})}\BibitemShut {NoStop}%
\bibitem [{\citenamefont {Kaubruegger~\emph{et~al.}}(2019)}]{vqe_sss1}%
  \BibitemOpen
  \bibfield  {author} {\bibinfo {author} {\bibfnamefont {R.}~\bibnamefont
  {Kaubruegger~\emph{et~al.}}},\ }\bibfield  {title} {\enquote {\bibinfo
  {title} {Variational spin-squeezing algorithms on programmable quantum
  sensors},}\ }\href {\doibase 10.1103/PhysRevLett.123.260505} {\bibfield
  {journal} {\bibinfo  {journal} {Phys. Rev. Lett.}\ }\textbf {\bibinfo
  {volume} {123}},\ \bibinfo {pages} {260505} (\bibinfo {year}
  {2019})}\BibitemShut {NoStop}%
\bibitem [{\citenamefont {Cerezo~\emph {et~al.}}(2021)}]{vq_1}%
  \BibitemOpen
  \bibfield  {author} {\bibinfo {author} {\bibfnamefont {M.}~\bibnamefont
  {Cerezo~\emph {et~al.}}},\ }\bibfield  {title} {\enquote {\bibinfo {title}
  {Variational quantum algorithms},}\ }\href {\doibase
  10.1038/s42254-021-00348-9} {\bibfield  {journal} {\bibinfo  {journal} {Nat.
  Rev. Phys.}\ }\textbf {\bibinfo {volume} {3}},\ \bibinfo {pages} {625--644}
  (\bibinfo {year} {2021})}\BibitemShut {NoStop}%
\bibitem [{\citenamefont {Kokail~\emph {et~al.}}(2019)}]{vq_2}%
  \BibitemOpen
  \bibfield  {author} {\bibinfo {author} {\bibfnamefont {C.}~\bibnamefont
  {Kokail~\emph {et~al.}}},\ }\bibfield  {title} {\enquote {\bibinfo {title}
  {Self-verifying variational quantum simulation of lattice models},}\ }\href
  {\doibase 10.1038/s41586-019-1177-4} {\bibfield  {journal} {\bibinfo
  {journal} {Nature}\ }\textbf {\bibinfo {volume} {569}},\ \bibinfo {pages}
  {355--360} (\bibinfo {year} {2019})}\BibitemShut {NoStop}%
\bibitem [{\citenamefont {McClean}\ \emph {et~al.}(2016)\citenamefont
  {McClean}, \citenamefont {Romero}, \citenamefont {Babbush},\ and\
  \citenamefont {Aspuru-Guzik}}]{vq_3}%
  \BibitemOpen
  \bibfield  {author} {\bibinfo {author} {\bibfnamefont {J.~R.}\ \bibnamefont
  {McClean}}, \bibinfo {author} {\bibfnamefont {J.}~\bibnamefont {Romero}},
  \bibinfo {author} {\bibfnamefont {R.}~\bibnamefont {Babbush}}, \ and\
  \bibinfo {author} {\bibfnamefont {A.}~\bibnamefont {Aspuru-Guzik}},\
  }\bibfield  {title} {\enquote {\bibinfo {title} {The theory of variational
  hybrid quantum-classical algorithms},}\ }\href {\doibase
  10.1088/1367-2630/18/2/023023} {\bibfield  {journal} {\bibinfo  {journal}
  {New J. Phys.}\ }\textbf {\bibinfo {volume} {18}},\ \bibinfo {pages} {023023}
  (\bibinfo {year} {2016})}\BibitemShut {NoStop}%
\bibitem [{\citenamefont {\emph {et~al.}}(2018)}]{vq_4}%
  \BibitemOpen
  \bibfield  {author} {\bibinfo {author} {\bibfnamefont {N.~Moll}\ \bibnamefont
  {\emph {et~al.}}},\ }\bibfield  {title} {\enquote {\bibinfo {title} {Quantum
  optimization using variational algorithms on near-term quantum devices},}\
  }\href {\doibase 10.1088/2058-9565/aab822} {\bibfield  {journal} {\bibinfo
  {journal} {Quantum Sci. Technol.}\ }\textbf {\bibinfo {volume} {3}},\
  \bibinfo {pages} {030503} (\bibinfo {year} {2018})}\BibitemShut {NoStop}%
\bibitem [{\citenamefont {Bharti~\emph {et~al.}}(2022)}]{vq_5}%
  \BibitemOpen
  \bibfield  {author} {\bibinfo {author} {\bibfnamefont {K.}~\bibnamefont
  {Bharti~\emph {et~al.}}},\ }\bibfield  {title} {\enquote {\bibinfo {title}
  {Noisy intermediate-scale quantum algorithms},}\ }\href {\doibase
  10.1103/RevModPhys.94.015004} {\bibfield  {journal} {\bibinfo  {journal}
  {Rev. Mod. Phys.}\ }\textbf {\bibinfo {volume} {94}},\ \bibinfo {pages}
  {015004} (\bibinfo {year} {2022})}\BibitemShut {NoStop}%
\bibitem [{\citenamefont {Huo}\ \emph {et~al.}(2022)\citenamefont {Huo},
  \citenamefont {Zhuang}, \citenamefont {Huang},\ and\ \citenamefont
  {Lee}}]{vqe_sss2}%
  \BibitemOpen
  \bibfield  {author} {\bibinfo {author} {\bibfnamefont {H.}~\bibnamefont
  {Huo}}, \bibinfo {author} {\bibfnamefont {M.}~\bibnamefont {Zhuang}},
  \bibinfo {author} {\bibfnamefont {J.}~\bibnamefont {Huang}}, \ and\ \bibinfo
  {author} {\bibfnamefont {C.}~\bibnamefont {Lee}},\ }\bibfield  {title}
  {\enquote {\bibinfo {title} {Machine optimized quantum metrology of
  concurrent entanglement generation and sensing},}\ }\href {\doibase
  10.1088/2058-9565/ac51af} {\bibfield  {journal} {\bibinfo  {journal} {Quantum
  Sci. Technol.}\ }\textbf {\bibinfo {volume} {7}},\ \bibinfo {pages} {025010}
  (\bibinfo {year} {2022})}\BibitemShut {NoStop}%
\bibitem [{\citenamefont {Zheng~\emph {et~al.}}(2022)}]{vqe_sss3}%
  \BibitemOpen
  \bibfield  {author} {\bibinfo {author} {\bibfnamefont {T.-X.}\ \bibnamefont
  {Zheng~\emph {et~al.}}},\ }\bibfield  {title} {\enquote {\bibinfo {title}
  {Preparation of metrological states in dipolar-interacting spin systems},}\
  }\href {\doibase 10.1038/s41534-022-00667-4} {\bibfield  {journal} {\bibinfo
  {journal} {npj Quantum Inf.}\ }\textbf {\bibinfo {volume} {8}},\ \bibinfo
  {pages} {150} (\bibinfo {year} {2022})}\BibitemShut {NoStop}%
\bibitem [{\citenamefont {{Santra}}\ \emph {et~al.}(2022)\citenamefont
  {{Santra}}, \citenamefont {{Jendrzejewski}}, \citenamefont {{Hauke}},\ and\
  \citenamefont {{Egger}}}]{vqe_sss4}%
  \BibitemOpen
  \bibfield  {author} {\bibinfo {author} {\bibfnamefont {G.~C.}\ \bibnamefont
  {{Santra}}}, \bibinfo {author} {\bibfnamefont {F.}~\bibnamefont
  {{Jendrzejewski}}}, \bibinfo {author} {\bibfnamefont {P.}~\bibnamefont
  {{Hauke}}}, \ and\ \bibinfo {author} {\bibfnamefont {D.~J.}\ \bibnamefont
  {{Egger}}},\ }\bibfield  {title} {\enquote {\bibinfo {title} {{Squeezing and
  quantum approximate optimization}},}\ }\href@noop {} {\bibfield  {journal}
  {\bibinfo  {journal} {arXiv e-prints}\ ,\ \bibinfo {eid} {arXiv:2205.10383}}
  (\bibinfo {year} {2022})},\ \Eprint {http://arxiv.org/abs/2205.10383}
  {arXiv:2205.10383 [quant-ph]} \BibitemShut {NoStop}%
\bibitem [{\citenamefont {Kaubruegger}\ \emph {et~al.}(2021)\citenamefont
  {Kaubruegger}, \citenamefont {Vasilyev}, \citenamefont {Schulte},
  \citenamefont {Hammerer},\ and\ \citenamefont {Zoller}}]{vqe_sss5}%
  \BibitemOpen
  \bibfield  {author} {\bibinfo {author} {\bibfnamefont {R.}~\bibnamefont
  {Kaubruegger}}, \bibinfo {author} {\bibfnamefont {D.~V.}\ \bibnamefont
  {Vasilyev}}, \bibinfo {author} {\bibfnamefont {M.}~\bibnamefont {Schulte}},
  \bibinfo {author} {\bibfnamefont {K.}~\bibnamefont {Hammerer}}, \ and\
  \bibinfo {author} {\bibfnamefont {P.}~\bibnamefont {Zoller}},\ }\bibfield
  {title} {\enquote {\bibinfo {title} {Quantum variational optimization of
  {R}amsey interferometry and atomic clocks},}\ }\href {\doibase
  10.1103/PhysRevX.11.041045} {\bibfield  {journal} {\bibinfo  {journal} {Phys.
  Rev. X}\ }\textbf {\bibinfo {volume} {11}},\ \bibinfo {pages} {041045}
  (\bibinfo {year} {2021})}\BibitemShut {NoStop}%
\bibitem [{\citenamefont {Marciniak~\emph {et~al.}}(2022)}]{vqe_sss6}%
  \BibitemOpen
  \bibfield  {author} {\bibinfo {author} {\bibfnamefont {C.~D.}\ \bibnamefont
  {Marciniak~\emph {et~al.}}},\ }\bibfield  {title} {\enquote {\bibinfo {title}
  {Optimal metrology with programmable quantum sensors},}\ }\href {\doibase
  10.1038/s41586-022-04435-4} {\bibfield  {journal} {\bibinfo  {journal}
  {Nature}\ }\textbf {\bibinfo {volume} {603}},\ \bibinfo {pages} {604--609}
  (\bibinfo {year} {2022})}\BibitemShut {NoStop}%
\bibitem [{\citenamefont {Gessner}\ \emph {et~al.}(2019)\citenamefont
  {Gessner}, \citenamefont {Smerzi},\ and\ \citenamefont
  {Pezz\`e}}]{PhysRevLett.122.090503}%
  \BibitemOpen
  \bibfield  {author} {\bibinfo {author} {\bibfnamefont {M.}~\bibnamefont
  {Gessner}}, \bibinfo {author} {\bibfnamefont {A.}~\bibnamefont {Smerzi}}, \
  and\ \bibinfo {author} {\bibfnamefont {L.}~\bibnamefont {Pezz\`e}},\
  }\bibfield  {title} {\enquote {\bibinfo {title} {Metrological nonlinear
  squeezing parameter},}\ }\href {\doibase 10.1103/PhysRevLett.122.090503}
  {\bibfield  {journal} {\bibinfo  {journal} {Phys. Rev. Lett.}\ }\textbf
  {\bibinfo {volume} {122}},\ \bibinfo {pages} {090503} (\bibinfo {year}
  {2019})}\BibitemShut {NoStop}%
\bibitem [{\citenamefont {Wu~\emph {et~al.}}(2021)}]{sc1}%
  \BibitemOpen
  \bibfield  {author} {\bibinfo {author} {\bibfnamefont {Y.}~\bibnamefont
  {Wu~\emph {et~al.}}},\ }\bibfield  {title} {\enquote {\bibinfo {title}
  {Strong quantum computational advantage using a superconducting quantum
  processor},}\ }\href {\doibase 10.1103/PhysRevLett.127.180501} {\bibfield
  {journal} {\bibinfo  {journal} {Phys. Rev. Lett.}\ }\textbf {\bibinfo
  {volume} {127}},\ \bibinfo {pages} {180501} (\bibinfo {year}
  {2021})}\BibitemShut {NoStop}%
\bibitem [{\citenamefont {Arute~\emph {et~al.}}(2019)}]{sc2}%
  \BibitemOpen
  \bibfield  {author} {\bibinfo {author} {\bibfnamefont {F.}~\bibnamefont
  {Arute~\emph {et~al.}}},\ }\bibfield  {title} {\enquote {\bibinfo {title}
  {Quantum supremacy using a programmable superconducting processor},}\ }\href
  {\doibase 10.1038/s41586-019-1666-5} {\bibfield  {journal} {\bibinfo
  {journal} {Nature}\ }\textbf {\bibinfo {volume} {574}},\ \bibinfo {pages}
  {505--510} (\bibinfo {year} {2019})}\BibitemShut {NoStop}%
\bibitem [{\citenamefont {Neill~\emph {et~al.}}(2021)}]{sc3}%
  \BibitemOpen
  \bibfield  {author} {\bibinfo {author} {\bibfnamefont {C.}~\bibnamefont
  {Neill~\emph {et~al.}}},\ }\bibfield  {title} {\enquote {\bibinfo {title}
  {Accurately computing the electronic properties of a quantum ring},}\ }\href
  {\doibase 10.1038/s41586-021-03576-2} {\bibfield  {journal} {\bibinfo
  {journal} {Nature}\ }\textbf {\bibinfo {volume} {594}},\ \bibinfo {pages}
  {508--512} (\bibinfo {year} {2021})}\BibitemShut {NoStop}%
\bibitem [{\citenamefont {Zhang~\emph {et~al.}}(2022)}]{sc4}%
  \BibitemOpen
  \bibfield  {author} {\bibinfo {author} {\bibfnamefont {X.}~\bibnamefont
  {Zhang~\emph {et~al.}}},\ }\bibfield  {title} {\enquote {\bibinfo {title}
  {Digital quantum simulation of {F}loquet symmetry-protected topological
  phases},}\ }\href {\doibase 10.1038/s41586-022-04854-3} {\bibfield  {journal}
  {\bibinfo  {journal} {Nature}\ }\textbf {\bibinfo {volume} {607}},\ \bibinfo
  {pages} {468--473} (\bibinfo {year} {2022})}\BibitemShut {NoStop}%
\bibitem [{\citenamefont {Frey}\ and\ \citenamefont {Rachel}(2022)}]{sc7}%
  \BibitemOpen
  \bibfield  {author} {\bibinfo {author} {\bibfnamefont {P.}~\bibnamefont
  {Frey}}\ and\ \bibinfo {author} {\bibfnamefont {S.}~\bibnamefont {Rachel}},\
  }\bibfield  {title} {\enquote {\bibinfo {title} {Realization of a discrete
  time crystal on 57 qubits of a quantum computer},}\ }\href {\doibase
  10.1126/sciadv.abm7652} {\bibfield  {journal} {\bibinfo  {journal} {Sci.
  Adv.}\ }\textbf {\bibinfo {volume} {8}},\ \bibinfo {pages} {eabm7652}
  (\bibinfo {year} {2022})}\BibitemShut {NoStop}%
\bibitem [{\citenamefont {Kandala}\ \emph {et~al.}(2017)\citenamefont
  {Kandala}, \citenamefont {Mezzacapo}, \citenamefont {Temme}, \citenamefont
  {Takita}, \citenamefont {Brink}, \citenamefont {Chow},\ and\ \citenamefont
  {Gambetta}}]{sc8}%
  \BibitemOpen
  \bibfield  {author} {\bibinfo {author} {\bibfnamefont {A.}~\bibnamefont
  {Kandala}}, \bibinfo {author} {\bibfnamefont {A.}~\bibnamefont {Mezzacapo}},
  \bibinfo {author} {\bibfnamefont {K.}~\bibnamefont {Temme}}, \bibinfo
  {author} {\bibfnamefont {M.}~\bibnamefont {Takita}}, \bibinfo {author}
  {\bibfnamefont {M.}~\bibnamefont {Brink}}, \bibinfo {author} {\bibfnamefont
  {J.~M.}\ \bibnamefont {Chow}}, \ and\ \bibinfo {author} {\bibfnamefont
  {J.~M.}\ \bibnamefont {Gambetta}},\ }\bibfield  {title} {\enquote {\bibinfo
  {title} {Hardware-efficient variational quantum eigensolver for small
  molecules and quantum magnets},}\ }\href {\doibase
  doi.org/10.1038/nature23879} {\bibfield  {journal} {\bibinfo  {journal}
  {Nature}\ }\textbf {\bibinfo {volume} {549}},\ \bibinfo {pages} {242--246}
  (\bibinfo {year} {2017})}\BibitemShut {NoStop}%
\bibitem [{\citenamefont {Bluvstein~\emph {et~al.}}(2022)}]{rydberg_gate1}%
  \BibitemOpen
  \bibfield  {author} {\bibinfo {author} {\bibfnamefont {D.}~\bibnamefont
  {Bluvstein~\emph {et~al.}}},\ }\bibfield  {title} {\enquote {\bibinfo {title}
  {A quantum processor based on coherent transport of entangled atom arrays},}\
  }\href {\doibase 10.1038/s41586-022-04592-6} {\bibfield  {journal} {\bibinfo
  {journal} {Nature}\ }\textbf {\bibinfo {volume} {604}},\ \bibinfo {pages}
  {451--456} (\bibinfo {year} {2022})}\BibitemShut {NoStop}%
\bibitem [{\citenamefont {Graham~\emph {et~al.}}(2022)}]{rydberg_gate2}%
  \BibitemOpen
  \bibfield  {author} {\bibinfo {author} {\bibfnamefont {T.~M.}\ \bibnamefont
  {Graham~\emph {et~al.}}},\ }\bibfield  {title} {\enquote {\bibinfo {title}
  {Multi-qubit entanglement and algorithms on a neutral-atom quantum
  computer},}\ }\href {\doibase 10.1038/s41586-022-04603-6} {\bibfield
  {journal} {\bibinfo  {journal} {Nature}\ }\textbf {\bibinfo {volume} {604}},\
  \bibinfo {pages} {457--462} (\bibinfo {year} {2022})}\BibitemShut {NoStop}%
\bibitem [{\citenamefont {Nakaji}\ and\ \citenamefont
  {Yamamoto}(2021)}]{Nakaji2021expressibilityof}%
  \BibitemOpen
  \bibfield  {author} {\bibinfo {author} {\bibfnamefont {K.}~\bibnamefont
  {Nakaji}}\ and\ \bibinfo {author} {\bibfnamefont {N.}~\bibnamefont
  {Yamamoto}},\ }\bibfield  {title} {\enquote {\bibinfo {title} {Expressibility
  of the alternating layered ansatz for quantum computation},}\ }\href
  {\doibase 10.22331/q-2021-04-19-434} {\bibfield  {journal} {\bibinfo
  {journal} {{Quantum}}\ }\textbf {\bibinfo {volume} {5}},\ \bibinfo {pages}
  {434} (\bibinfo {year} {2021})}\BibitemShut {NoStop}%
\bibitem [{\citenamefont {Cerezo}\ \emph {et~al.}(2021)\citenamefont {Cerezo},
  \citenamefont {Sone}, \citenamefont {Volkoff}, \citenamefont {Cincio},\ and\
  \citenamefont {Coles}}]{tat_expr1}%
  \BibitemOpen
  \bibfield  {author} {\bibinfo {author} {\bibfnamefont {M.}~\bibnamefont
  {Cerezo}}, \bibinfo {author} {\bibfnamefont {A.}~\bibnamefont {Sone}},
  \bibinfo {author} {\bibfnamefont {T.}~\bibnamefont {Volkoff}}, \bibinfo
  {author} {\bibfnamefont {L.}~\bibnamefont {Cincio}}, \ and\ \bibinfo {author}
  {\bibfnamefont {P.~J.}\ \bibnamefont {Coles}},\ }\bibfield  {title} {\enquote
  {\bibinfo {title} {Cost function dependent barren plateaus in shallow
  parametrized quantum circuits},}\ }\href {\doibase
  10.1038/s41467-021-21728-w} {\bibfield  {journal} {\bibinfo  {journal} {Nat.
  Commun.}\ }\textbf {\bibinfo {volume} {12}},\ \bibinfo {pages} {1791}
  (\bibinfo {year} {2021})}\BibitemShut {NoStop}%
\bibitem [{\citenamefont {Kivlichan}\ \emph {et~al.}(2018)\citenamefont
  {Kivlichan}, \citenamefont {McClean}, \citenamefont {Wiebe}, \citenamefont
  {Gidney}, \citenamefont {Aspuru-Guzik}, \citenamefont {Chan},\ and\
  \citenamefont {Babbush}}]{PhysRevLett.120.110501}%
  \BibitemOpen
  \bibfield  {author} {\bibinfo {author} {\bibfnamefont {I.~D.}\ \bibnamefont
  {Kivlichan}}, \bibinfo {author} {\bibfnamefont {J.}~\bibnamefont {McClean}},
  \bibinfo {author} {\bibfnamefont {N.}~\bibnamefont {Wiebe}}, \bibinfo
  {author} {\bibfnamefont {C.}~\bibnamefont {Gidney}}, \bibinfo {author}
  {\bibfnamefont {A.}~\bibnamefont {Aspuru-Guzik}}, \bibinfo {author}
  {\bibfnamefont {G.~K.-L.}\ \bibnamefont {Chan}}, \ and\ \bibinfo {author}
  {\bibfnamefont {R.}~\bibnamefont {Babbush}},\ }\bibfield  {title} {\enquote
  {\bibinfo {title} {Quantum simulation of electronic structure with linear
  depth and connectivity},}\ }\href {\doibase 10.1103/PhysRevLett.120.110501}
  {\bibfield  {journal} {\bibinfo  {journal} {Phys. Rev. Lett.}\ }\textbf
  {\bibinfo {volume} {120}},\ \bibinfo {pages} {110501} (\bibinfo {year}
  {2018})}\BibitemShut {NoStop}%
\bibitem [{\citenamefont {Foxen~\emph
  {et~al.}}(2020)}]{PhysRevLett.125.120504}%
  \BibitemOpen
  \bibfield  {author} {\bibinfo {author} {\bibfnamefont {B.}~\bibnamefont
  {Foxen~\emph {et~al.}}} (\bibinfo {collaboration} {Google AI Quantum}),\
  }\bibfield  {title} {\enquote {\bibinfo {title} {Demonstrating a continuous
  set of two-qubit gates for near-term quantum algorithms},}\ }\href {\doibase
  10.1103/PhysRevLett.125.120504} {\bibfield  {journal} {\bibinfo  {journal}
  {Phys. Rev. Lett.}\ }\textbf {\bibinfo {volume} {125}},\ \bibinfo {pages}
  {120504} (\bibinfo {year} {2020})}\BibitemShut {NoStop}%
\bibitem [{\citenamefont {Sim}\ \emph {et~al.}(2019)\citenamefont {Sim},
  \citenamefont {Johnson},\ and\ \citenamefont {Aspuru-Guzik}}]{expr_add}%
  \BibitemOpen
  \bibfield  {author} {\bibinfo {author} {\bibfnamefont {S.}~\bibnamefont
  {Sim}}, \bibinfo {author} {\bibfnamefont {P.~D.}\ \bibnamefont {Johnson}}, \
  and\ \bibinfo {author} {\bibfnamefont {A.}~\bibnamefont {Aspuru-Guzik}},\
  }\bibfield  {title} {\enquote {\bibinfo {title} {Expressibility and
  entangling capability of parameterized quantum circuits for hybrid
  quantum-classical algorithms},}\ }\href {\doibase
  https://doi.org/10.1002/qute.201900070} {\bibfield  {journal} {\bibinfo
  {journal} {Adv. Quantum Technol.}\ }\textbf {\bibinfo {volume} {2}},\
  \bibinfo {pages} {1900070} (\bibinfo {year} {2019})}\BibitemShut {NoStop}%
\bibitem [{\citenamefont {Weidenfeller}\ \emph {et~al.}(2022)\citenamefont
  {Weidenfeller}, \citenamefont {Valor}, \citenamefont {Gacon}, \citenamefont
  {Tornow}, \citenamefont {Bello}, \citenamefont {Woerner},\ and\ \citenamefont
  {Egger}}]{Weidenfeller2022scalingofquantum}%
  \BibitemOpen
  \bibfield  {author} {\bibinfo {author} {\bibfnamefont {J.}~\bibnamefont
  {Weidenfeller}}, \bibinfo {author} {\bibfnamefont {L.~C.}\ \bibnamefont
  {Valor}}, \bibinfo {author} {\bibfnamefont {J.}~\bibnamefont {Gacon}},
  \bibinfo {author} {\bibfnamefont {C.}~\bibnamefont {Tornow}}, \bibinfo
  {author} {\bibfnamefont {L.}~\bibnamefont {Bello}}, \bibinfo {author}
  {\bibfnamefont {S.}~\bibnamefont {Woerner}}, \ and\ \bibinfo {author}
  {\bibfnamefont {D.~J.}\ \bibnamefont {Egger}},\ }\bibfield  {title} {\enquote
  {\bibinfo {title} {Scaling of the quantum approximate optimization algorithm
  on superconducting qubit based hardware},}\ }\href {\doibase
  10.22331/q-2022-12-07-870} {\bibfield  {journal} {\bibinfo  {journal}
  {{Quantum}}\ }\textbf {\bibinfo {volume} {6}},\ \bibinfo {pages} {870}
  (\bibinfo {year} {2022})}\BibitemShut {NoStop}%
\bibitem [{\citenamefont {Harrigan~\emph{et~al.}}(2021)}]{Harrigan:2021we}%
  \BibitemOpen
  \bibfield  {author} {\bibinfo {author} {\bibfnamefont {M.~P.}\ \bibnamefont
  {Harrigan~\emph{et~al.}}},\ }\bibfield  {title} {\enquote {\bibinfo {title}
  {Quantum approximate optimization of non-planar graph problems on a planar
  superconducting processor},}\ }\href {\doibase 10.1038/s41567-020-01105-y}
  {\bibfield  {journal} {\bibinfo  {journal} {Nature Physics}\ }\textbf
  {\bibinfo {volume} {17}},\ \bibinfo {pages} {332--336} (\bibinfo {year}
  {2021})}\BibitemShut {NoStop}%
\bibitem [{\citenamefont {Barends~\emph
  {et~al.}}(2019)}]{PhysRevLett.123.210501}%
  \BibitemOpen
  \bibfield  {author} {\bibinfo {author} {\bibfnamefont {R.}~\bibnamefont
  {Barends~\emph {et~al.}}},\ }\bibfield  {title} {\enquote {\bibinfo {title}
  {Diabatic gates for frequency-tunable superconducting qubits},}\ }\href
  {\doibase 10.1103/PhysRevLett.123.210501} {\bibfield  {journal} {\bibinfo
  {journal} {Phys. Rev. Lett.}\ }\textbf {\bibinfo {volume} {123}},\ \bibinfo
  {pages} {210501} (\bibinfo {year} {2019})}\BibitemShut {NoStop}%
\bibitem [{\citenamefont {Kockum}\ and\ \citenamefont
  {Nori}(2019)}]{Kockum2019}%
  \BibitemOpen
  \bibfield  {author} {\bibinfo {author} {\bibfnamefont {A.~F.}\ \bibnamefont
  {Kockum}}\ and\ \bibinfo {author} {\bibfnamefont {F.}~\bibnamefont {Nori}},\
  }\enquote {\bibinfo {title} {Quantum bits with {J}osephson junctions},}\ in\
  \href {\doibase 10.1007/978-3-030-20726-7_17} {\emph {\bibinfo {booktitle}
  {Fundamentals and Frontiers of the Josephson Effect}}},\ \bibinfo {editor}
  {edited by\ \bibinfo {editor} {\bibfnamefont {Francesco}\ \bibnamefont
  {Tafuri}}}\ (\bibinfo  {publisher} {Springer International Publishing},\
  \bibinfo {address} {Cham},\ \bibinfo {year} {2019})\ pp.\ \bibinfo {pages}
  {703--741}\BibitemShut {NoStop}%
\bibitem [{\citenamefont {Chen~\emph {et~al.}}(2021)}]{sc5}%
  \BibitemOpen
  \bibfield  {author} {\bibinfo {author} {\bibfnamefont {F.}~\bibnamefont
  {Chen~\emph {et~al.}}},\ }\bibfield  {title} {\enquote {\bibinfo {title}
  {Observation of strong and weak thermalization in a superconducting quantum
  processor},}\ }\href {\doibase 10.1103/PhysRevLett.127.020602} {\bibfield
  {journal} {\bibinfo  {journal} {Phys. Rev. Lett.}\ }\textbf {\bibinfo
  {volume} {127}},\ \bibinfo {pages} {020602} (\bibinfo {year}
  {2021})}\BibitemShut {NoStop}%
\bibitem [{\citenamefont {Zhu~\emph {et~al.}}(2022)}]{sc6}%
  \BibitemOpen
  \bibfield  {author} {\bibinfo {author} {\bibfnamefont {Q.}~\bibnamefont
  {Zhu~\emph {et~al.}}},\ }\bibfield  {title} {\enquote {\bibinfo {title}
  {Observation of thermalization and information scrambling in a
  superconducting quantum processor},}\ }\href {\doibase
  10.1103/PhysRevLett.128.160502} {\bibfield  {journal} {\bibinfo  {journal}
  {Phys. Rev. Lett.}\ }\textbf {\bibinfo {volume} {128}},\ \bibinfo {pages}
  {160502} (\bibinfo {year} {2022})}\BibitemShut {NoStop}%
\bibitem [{\citenamefont {Yan~\emph {et~al.}}(2019)}]{sc9}%
  \BibitemOpen
  \bibfield  {author} {\bibinfo {author} {\bibfnamefont {Z.}~\bibnamefont
  {Yan~\emph {et~al.}}},\ }\bibfield  {title} {\enquote {\bibinfo {title}
  {Strongly correlated quantum walks with a 12-qubit superconducting
  processor.}}\ }\href {\doibase 10.1126/science.aaw1611} {\bibfield  {journal}
  {\bibinfo  {journal} {Science}\ }\textbf {\bibinfo {volume} {364}},\ \bibinfo
  {pages} {753--756} (\bibinfo {year} {2019})}\BibitemShut {NoStop}%
\bibitem [{\citenamefont {Roushan~\emph {et~al.}}(2017)}]{sc10}%
  \BibitemOpen
  \bibfield  {author} {\bibinfo {author} {\bibfnamefont {P.}~\bibnamefont
  {Roushan~\emph {et~al.}}},\ }\bibfield  {title} {\enquote {\bibinfo {title}
  {Spectroscopic signatures of localization with interacting photons in
  superconducting qubits.}}\ }\href {\doibase 10.1126/science.aao1401}
  {\bibfield  {journal} {\bibinfo  {journal} {Science}\ }\textbf {\bibinfo
  {volume} {358}},\ \bibinfo {pages} {1175--1179} (\bibinfo {year}
  {2017})}\BibitemShut {NoStop}%
\bibitem [{\citenamefont {Kannan~\emph {et~al.}}(2020)}]{Kannan:2020uv}%
  \BibitemOpen
  \bibfield  {author} {\bibinfo {author} {\bibfnamefont {B.}~\bibnamefont
  {Kannan~\emph {et~al.}}},\ }\bibfield  {title} {\enquote {\bibinfo {title}
  {Waveguide quantum electrodynamics with superconducting artificial giant
  atoms},}\ }\href {\doibase 10.1038/s41586-020-2529-9} {\bibfield  {journal}
  {\bibinfo  {journal} {Nature}\ }\textbf {\bibinfo {volume} {583}},\ \bibinfo
  {pages} {775--779} (\bibinfo {year} {2020})}\BibitemShut {NoStop}%
\bibitem [{\citenamefont {You~\emph{et~al.}}(2007)}]{PhysRevB.75.140515}%
  \BibitemOpen
  \bibfield  {author} {\bibinfo {author} {\bibfnamefont {J.~Q.}\ \bibnamefont
  {You~\emph{et~al.}}},\ }\bibfield  {title} {\enquote {\bibinfo {title}
  {Low-decoherence flux qubit},}\ }\href {\doibase 10.1103/PhysRevB.75.140515}
  {\bibfield  {journal} {\bibinfo  {journal} {Phys. Rev. B}\ }\textbf {\bibinfo
  {volume} {75}},\ \bibinfo {pages} {140515} (\bibinfo {year}
  {2007})}\BibitemShut {NoStop}%
\bibitem [{\citenamefont {Gu~\emph{et~al.}}(2017)}]{Gu:2017ut}%
  \BibitemOpen
  \bibfield  {author} {\bibinfo {author} {\bibfnamefont {X.}~\bibnamefont
  {Gu~\emph{et~al.}}},\ }\bibfield  {title} {\enquote {\bibinfo {title}
  {Microwave photonics with superconducting quantum circuits},}\ }\href
  {\doibase https://doi.org/10.1016/j.physrep.2017.10.002} {\bibfield
  {journal} {\bibinfo  {journal} {Phys. Rep.}\ }\textbf {\bibinfo {volume}
  {718-719}},\ \bibinfo {pages} {1--102} (\bibinfo {year} {2017})}\BibitemShut
  {NoStop}%
\bibitem [{\citenamefont {Parra-Rodriguez}\ \emph {et~al.}(2020)\citenamefont
  {Parra-Rodriguez}, \citenamefont {Lougovski}, \citenamefont {Lamata},
  \citenamefont {Solano},\ and\ \citenamefont {Sanz}}]{PhysRevA.101.022305}%
  \BibitemOpen
  \bibfield  {author} {\bibinfo {author} {\bibfnamefont {A.}~\bibnamefont
  {Parra-Rodriguez}}, \bibinfo {author} {\bibfnamefont {P.}~\bibnamefont
  {Lougovski}}, \bibinfo {author} {\bibfnamefont {L.}~\bibnamefont {Lamata}},
  \bibinfo {author} {\bibfnamefont {E.}~\bibnamefont {Solano}}, \ and\ \bibinfo
  {author} {\bibfnamefont {M.}~\bibnamefont {Sanz}},\ }\bibfield  {title}
  {\enquote {\bibinfo {title} {Digital-analog quantum computation},}\ }\href
  {\doibase 10.1103/PhysRevA.101.022305} {\bibfield  {journal} {\bibinfo
  {journal} {Phys. Rev. A}\ }\textbf {\bibinfo {volume} {101}},\ \bibinfo
  {pages} {022305} (\bibinfo {year} {2020})}\BibitemShut {NoStop}%
\bibitem [{\citenamefont {Mi~\emph {et~al.}}(2021)}]{coherent_error_OTOC}%
  \BibitemOpen
  \bibfield  {author} {\bibinfo {author} {\bibfnamefont {X.}~\bibnamefont
  {Mi~\emph {et~al.}}},\ }\bibfield  {title} {\enquote {\bibinfo {title}
  {Information scrambling in quantum circuits.}}\ }\href {\doibase
  10.1126/science.abg5029} {\bibfield  {journal} {\bibinfo  {journal}
  {Science}\ }\textbf {\bibinfo {volume} {374}},\ \bibinfo {pages} {1479--1483}
  (\bibinfo {year} {2021})}\BibitemShut {NoStop}%
\bibitem [{\citenamefont {Degen}\ \emph {et~al.}(2017)\citenamefont {Degen},
  \citenamefont {Reinhard},\ and\ \citenamefont
  {Cappellaro}}]{RevModPhys.89.035002}%
  \BibitemOpen
  \bibfield  {author} {\bibinfo {author} {\bibfnamefont {C.~L.}\ \bibnamefont
  {Degen}}, \bibinfo {author} {\bibfnamefont {F.}~\bibnamefont {Reinhard}}, \
  and\ \bibinfo {author} {\bibfnamefont {P.}~\bibnamefont {Cappellaro}},\
  }\bibfield  {title} {\enquote {\bibinfo {title} {Quantum sensing},}\ }\href
  {\doibase 10.1103/RevModPhys.89.035002} {\bibfield  {journal} {\bibinfo
  {journal} {Rev. Mod. Phys.}\ }\textbf {\bibinfo {volume} {89}},\ \bibinfo
  {pages} {035002} (\bibinfo {year} {2017})}\BibitemShut {NoStop}%
\bibitem [{\citenamefont {Schollw\"ock}(2005)}]{RevModPhys.77.259}%
  \BibitemOpen
  \bibfield  {author} {\bibinfo {author} {\bibfnamefont {U.}~\bibnamefont
  {Schollw\"ock}},\ }\bibfield  {title} {\enquote {\bibinfo {title} {The
  density-matrix renormalization group},}\ }\href {\doibase
  10.1103/RevModPhys.77.259} {\bibfield  {journal} {\bibinfo  {journal} {Rev.
  Mod. Phys.}\ }\textbf {\bibinfo {volume} {77}},\ \bibinfo {pages} {259--315}
  (\bibinfo {year} {2005})}\BibitemShut {NoStop}%
\bibitem [{\citenamefont {Henriet}\ \emph {et~al.}(2020)\citenamefont
  {Henriet}, \citenamefont {Beguin}, \citenamefont {Signoles}, \citenamefont
  {Lahaye}, \citenamefont {Browaeys}, \citenamefont {Reymond},\ and\
  \citenamefont {Jurczak}}]{Henriet2020quantumcomputing}%
  \BibitemOpen
  \bibfield  {author} {\bibinfo {author} {\bibfnamefont {L.}~\bibnamefont
  {Henriet}}, \bibinfo {author} {\bibfnamefont {L.}~\bibnamefont {Beguin}},
  \bibinfo {author} {\bibfnamefont {A.}~\bibnamefont {Signoles}}, \bibinfo
  {author} {\bibfnamefont {T.}~\bibnamefont {Lahaye}}, \bibinfo {author}
  {\bibfnamefont {A.}~\bibnamefont {Browaeys}}, \bibinfo {author}
  {\bibfnamefont {G.-O.}\ \bibnamefont {Reymond}}, \ and\ \bibinfo {author}
  {\bibfnamefont {C.}~\bibnamefont {Jurczak}},\ }\bibfield  {title} {\enquote
  {\bibinfo {title} {Quantum computing with neutral atoms},}\ }\href {\doibase
  10.22331/q-2020-09-21-327} {\bibfield  {journal} {\bibinfo  {journal}
  {{Quantum}}\ }\textbf {\bibinfo {volume} {4}},\ \bibinfo {pages} {327}
  (\bibinfo {year} {2020})}\BibitemShut {NoStop}%
\bibitem [{\citenamefont {Browaeys}\ and\ \citenamefont
  {Lahaye}(2020)}]{Browaeys:2020tl}%
  \BibitemOpen
  \bibfield  {author} {\bibinfo {author} {\bibfnamefont {A.}~\bibnamefont
  {Browaeys}}\ and\ \bibinfo {author} {\bibfnamefont {T.}~\bibnamefont
  {Lahaye}},\ }\bibfield  {title} {\enquote {\bibinfo {title} {Many-body
  physics with individually controlled {R}ydberg atoms},}\ }\href {\doibase
  10.1038/s41567-019-0733-z} {\bibfield  {journal} {\bibinfo  {journal} {Nat.
  Phys.}\ }\textbf {\bibinfo {volume} {16}},\ \bibinfo {pages} {132--142}
  (\bibinfo {year} {2020})}\BibitemShut {NoStop}%
\bibitem [{\citenamefont {Zeiher}\ \emph {et~al.}(2017)\citenamefont {Zeiher},
  \citenamefont {Choi}, \citenamefont {Rubio-Abadal}, \citenamefont {Pohl},
  \citenamefont {van Bijnen}, \citenamefont {Bloch},\ and\ \citenamefont
  {Gross}}]{PhysRevX.7.041063}%
  \BibitemOpen
  \bibfield  {author} {\bibinfo {author} {\bibfnamefont {J.}~\bibnamefont
  {Zeiher}}, \bibinfo {author} {\bibfnamefont {J.-y.}\ \bibnamefont {Choi}},
  \bibinfo {author} {\bibfnamefont {A.}~\bibnamefont {Rubio-Abadal}}, \bibinfo
  {author} {\bibfnamefont {T.}~\bibnamefont {Pohl}}, \bibinfo {author}
  {\bibfnamefont {R.}~\bibnamefont {van Bijnen}}, \bibinfo {author}
  {\bibfnamefont {I.}~\bibnamefont {Bloch}}, \ and\ \bibinfo {author}
  {\bibfnamefont {C.}~\bibnamefont {Gross}},\ }\bibfield  {title} {\enquote
  {\bibinfo {title} {Coherent many-body spin dynamics in a long-range
  interacting {I}sing chain},}\ }\href {\doibase 10.1103/PhysRevX.7.041063}
  {\bibfield  {journal} {\bibinfo  {journal} {Phys. Rev. X}\ }\textbf {\bibinfo
  {volume} {7}},\ \bibinfo {pages} {041063} (\bibinfo {year}
  {2017})}\BibitemShut {NoStop}%
\bibitem [{\citenamefont {Borish}\ \emph {et~al.}(2020)\citenamefont {Borish},
  \citenamefont {Markovi\ifmmode~\acute{c}\else \'{c}\fi{}}, \citenamefont
  {Hines}, \citenamefont {Rajagopal},\ and\ \citenamefont
  {Schleier-Smith}}]{PhysRevLett.124.063601}%
  \BibitemOpen
  \bibfield  {author} {\bibinfo {author} {\bibfnamefont {V.}~\bibnamefont
  {Borish}}, \bibinfo {author} {\bibfnamefont {O.}~\bibnamefont
  {Markovi\ifmmode~\acute{c}\else \'{c}\fi{}}}, \bibinfo {author}
  {\bibfnamefont {J.~A.}\ \bibnamefont {Hines}}, \bibinfo {author}
  {\bibfnamefont {S.~V.}\ \bibnamefont {Rajagopal}}, \ and\ \bibinfo {author}
  {\bibfnamefont {M.}~\bibnamefont {Schleier-Smith}},\ }\bibfield  {title}
  {\enquote {\bibinfo {title} {Transverse-field {I}sing dynamics in a
  {R}ydberg-dressed atomic gas},}\ }\href {\doibase
  10.1103/PhysRevLett.124.063601} {\bibfield  {journal} {\bibinfo  {journal}
  {Phys. Rev. Lett.}\ }\textbf {\bibinfo {volume} {124}},\ \bibinfo {pages}
  {063601} (\bibinfo {year} {2020})}\BibitemShut {NoStop}%
\bibitem [{\citenamefont {Ballester}\ \emph {et~al.}(2012)\citenamefont
  {Ballester}, \citenamefont {Romero}, \citenamefont {Garc\'{\i}a-Ripoll},
  \citenamefont {Deppe},\ and\ \citenamefont {Solano}}]{PhysRevX.2.021007}%
  \BibitemOpen
  \bibfield  {author} {\bibinfo {author} {\bibfnamefont {D.}~\bibnamefont
  {Ballester}}, \bibinfo {author} {\bibfnamefont {G.}~\bibnamefont {Romero}},
  \bibinfo {author} {\bibfnamefont {J.~J.}\ \bibnamefont {Garc\'{\i}a-Ripoll}},
  \bibinfo {author} {\bibfnamefont {F.}~\bibnamefont {Deppe}}, \ and\ \bibinfo
  {author} {\bibfnamefont {E.}~\bibnamefont {Solano}},\ }\bibfield  {title}
  {\enquote {\bibinfo {title} {Quantum simulation of the ultrastrong-coupling
  dynamics in circuit quantum electrodynamics},}\ }\href {\doibase
  10.1103/PhysRevX.2.021007} {\bibfield  {journal} {\bibinfo  {journal} {Phys.
  Rev. X}\ }\textbf {\bibinfo {volume} {2}},\ \bibinfo {pages} {021007}
  (\bibinfo {year} {2012})}\BibitemShut {NoStop}%
\bibitem [{\citenamefont {\emph{et~al.}}(2020)}]{Qin4868}%
  \BibitemOpen
  \bibfield  {author} {\bibinfo {author} {\bibfnamefont {W.~Qin}\ \bibnamefont
  {\emph{et~al.}}},\ }\bibfield  {title} {\enquote {\bibinfo {title} {Strong
  spin squeezing induced by weak squeezing of light inside a cavity},}\ }\href
  {\doibase doi:10.1515/nanoph-2020-0513} {\bibfield  {journal} {\bibinfo
  {journal} {Nanophotonics}\ }\textbf {\bibinfo {volume} {9}},\ \bibinfo
  {pages} {4853--4868} (\bibinfo {year} {2020})}\BibitemShut {NoStop}%
\bibitem [{\citenamefont
  {Macr\`{\i}~\emph{et~al.}}(2020)}]{PhysRevA.101.053818}%
  \BibitemOpen
  \bibfield  {author} {\bibinfo {author} {\bibfnamefont {V.}~\bibnamefont
  {Macr\`{\i}~\emph{et~al.}}},\ }\bibfield  {title} {\enquote {\bibinfo {title}
  {Spin squeezing by one-photon--two-atom excitation processes in atomic
  ensembles},}\ }\href {\doibase 10.1103/PhysRevA.101.053818} {\bibfield
  {journal} {\bibinfo  {journal} {Phys. Rev. A}\ }\textbf {\bibinfo {volume}
  {101}},\ \bibinfo {pages} {053818} (\bibinfo {year} {2020})}\BibitemShut
  {NoStop}%
\bibitem [{\citenamefont {Zanardi}\ \emph {et~al.}(2000)\citenamefont
  {Zanardi}, \citenamefont {Zalka},\ and\ \citenamefont
  {Faoro}}]{PhysRevA.62.030301}%
  \BibitemOpen
  \bibfield  {author} {\bibinfo {author} {\bibfnamefont {P.}~\bibnamefont
  {Zanardi}}, \bibinfo {author} {\bibfnamefont {C.}~\bibnamefont {Zalka}}, \
  and\ \bibinfo {author} {\bibfnamefont {L.}~\bibnamefont {Faoro}},\ }\bibfield
   {title} {\enquote {\bibinfo {title} {Entangling power of quantum
  evolutions},}\ }\href {\doibase 10.1103/PhysRevA.62.030301} {\bibfield
  {journal} {\bibinfo  {journal} {Phys. Rev. A}\ }\textbf {\bibinfo {volume}
  {62}},\ \bibinfo {pages} {030301} (\bibinfo {year} {2000})}\BibitemShut
  {NoStop}%
\end{thebibliography}%

\end{document}